\documentclass[runningheads]{llncs}
\usepackage{graphicx}
\usepackage{algorithm}
\usepackage{algorithm}
\usepackage[noend]{algpseudocode}
\usepackage{pgfplots}
\usepackage{cite}
\usepackage{amsmath,amssymb,amsfonts}
\usepackage{textcomp}
\usepackage{xcolor,colortbl}
\usepackage{comment}
\usepackage{booktabs}
\usepackage{multirow}
\graphicspath{{figures/}}
\usepackage{todonotes}
\usepackage{wrapfig}
\usepackage{lipsum}
\usepackage[caption=false]{subfig}
\usepackage[title]{appendix}
\usepackage{diagbox}

\pgfplotsset{compat=1.13,           
            width=\columnwidth,     
            height=0.75\columnwidth 
            }
            
\def\BibTeX{{\rm B\kern-.05em{\sc i\kern-.025em b}\kern-.08em
    T\kern-.1667em\lower.7ex\hbox{E}\kern-.125emX}}
    
\newcommand{\squishlist}{
	\begin{list}{$\bullet$}
		{ \setlength{\itemsep}{2pt}    \setlength{\parsep}{0pt}
			\setlength{\topsep}{5pt}     \setlength{\partopsep}{0pt}
			\setlength{\leftmargin}{1.35em} \setlength{\labelwidth}{1em}
			\setlength{\labelsep}{0.5em} } }
	
\newcommand{\squishend}{
	\end{list}  }

\newcolumntype{?}{!{\vrule width 2pt}}

\algnewcommand{\IIf}[1]{\State\algorithmicif\ #1\ \algorithmicthen}
\algnewcommand{\EElse}[1]{\State\algorithmicelse\ #1\ }
\algnewcommand{\EndIIf}{\unskip}
\algrenewcommand\algorithmicindent{1.2em}%

\begin{document}

\title{DeepFault: Fault Localization\\for Deep Neural Networks\thanks{This research was supported in part by Bogazici University; Research Fund 13662.}
\vspace*{-5mm}}

\author{Hasan Ferit Eniser\inst{1}\orcidID{0000-0002-3259-8794} \and
Simos Gerasimou\inst{2}\orcidID{0000-0002-2706-5272} \and
Alper Sen\inst{1}\orcidID{0000-0002-5508-6484}}
\authorrunning{H.F.Eniser et al.}
%
\institute{Bogazici University, Istanbul, Turkey \\ \email{\{hasan.eniser,alper.sen\}@boun.edu.tr} \and
University of York, York, UK \\
\email{simos.gerasimou@york.ac.uk}\\
}

\maketitle

\begin{abstract}
\vspace*{-7mm}
Deep Neural Networks (DNNs) are increasingly deployed in safety-critical 
applications including autonomous vehicles and medical diagnostics. 
To reduce the residual risk for unexpected DNN behaviour and provide evidence 
for their trustworthy operation, DNNs should be thoroughly tested. 
The DeepFault whitebox DNN testing approach presented in our paper addresses 
this challenge by employing suspiciousness measures inspired by fault 
localization to establish the hit spectrum of neurons and identify suspicious 
neurons whose weights have not been calibrated correctly and thus are considered
responsible for inadequate DNN performance.
DeepFault also uses a suspiciousness-guided algorithm to synthesize new inputs, 
from correctly classified inputs, that increase the activation values of 
suspicious neurons. 
Our empirical evaluation on several DNN instances trained on MNIST and 
CIFAR-10 datasets shows that DeepFault is effective in identifying suspicious 
 neurons. 
Also, the inputs synthesized by DeepFault closely resemble the original inputs, 
exercise the identified suspicious neurons and are highly adversarial.
\vspace*{-4mm}
\end{abstract}

\keywords{Deep Neural Networks \and Fault Localization \and Test Input 
Generation
\vspace*{-8mm}}
\section{Introduction}
\vspace*{-1em}

Deep Neural Networks (DNNs)~\cite{LecunBH2015} have demonstrated human-level 
capabilities in several intractable  machine learning tasks including image 
classification~\cite{CiresanMS2012}, natural language 
processing~\cite{SutskeverVQ2014} and speech recognition~\cite{HintonDYD2012}. 
These impressive achievements raised the expectations for
deploying DNNs in real-world applications, especially in 
safety-critical domains. Early-stage applications include 
air traffic control~\cite{julian2016policy}, 
medical diagnostics~\cite{litjens2017survey} and autonomous 
vehicles~\cite{bojarski2016end}. The responsibilities of DNNs in these 
applications vary from 
carrying out well-defined tasks (e.g., detecting abnormal network 
activity~\cite{CuiXCCWC2018}) to controlling the entire behaviour system  
(e.g., end-to-end learning in autonomous vehicles~\cite{bojarski2016end}).

Despite the anticipated benefits from a widespread adoption of DNNs, 
their deployment in safety-critical systems must be characterized by a high degree of 
dependability. 
Deviations from the expected behaviour or correct operation, as expected in 
safety-critical domains, can endanger human lives or cause significant 
financial loss. Arguably, DNN-based systems should be granted permission for 
use in the public domain only 
after exhibiting high levels of trustworthiness~\cite{BurtonGH2017}.

Software testing is the de facto instrument for analysing and evaluating the 
quality of a software system~\cite{jorgensen2013software}. Testing enables at 
one hand  to reduce the risk by proactively finding and eliminating problems 
(\textit{bugs}), and on the other hand to evidence, through using the testing 
results, that the system actually achieves the required levels of safety. 
Research contributions and advice on best practices for testing conventional 
software systems are plentiful;~\cite{wong2016survey}, for instance, 
provides a comprehensive review of the state-of-the-art testing approaches.

Nevertheless, there are significant challenges in applying traditional software 
testing techniques for assessing the quality of DNN-based software~\cite{SHK18}.
Most importantly, the little correlation between the behaviour 
of a DNN and the software used for its implementation means that the behaviour 
of the DNN cannot be explicitly encoded in the control flow structures of the 
software~\cite{salay2018analysis}. 
Furthermore, DNNs have very complex architectures, typically comprising 
thousand or millions of parameters, making it difficult, if not impossible, to 
determine a parameter's contribution to achieving a task. 
Likewise, since the behaviour of a DNN is heavily influenced by the data used 
during training, collecting enough data that enables exercising all potential 
DNN behaviour under all possible scenarios becomes a very challenging task.
Hence, there is a need for systematic and effective testing frameworks for 
evaluating the quality of DNN-based software~\cite{BurtonGH2017}. 

Recent research in the DNN testing area introduces novel white-box and 
black-box techniques 
for testing 
DNNs~\cite{pei2017deepxplore,Huang18,KFY18,MZSXJ18,MZXLLZW18,SHK18,SWRHKK18}.
Some techniques transform valid training data into adversarial through 
mutation-based heuristics~\cite{WWRHK18}, apply symbolic 
execution~\cite{GWZPK18}, combinatorial~\cite{MZXLLZW18} or concolic 
testing~\cite{SWRHKK18},  
while others propose new DNN-specific coverage criteria, e.g., neuron 
coverage~\cite{pei2017deepxplore} and its variants~\cite{MJX18} or 
MC/DC-inspired criteria~\cite{SDDFGK18}.  
We review related work in Section~\ref{sec:relatedWork}. 
These recent advances provide evidence that, while traditional software testing 
techniques are not directly applicable to testing DNNs, the sophisticated 
concepts and principles behind these techniques, if adapted appropriately, 
could be useful to the machine learning domain. 
Nevertheless, none of the proposed techniques uses \textit{fault 
localisation}~\cite{wong2016survey,pearson2017evaluating,artzi2010directed}, 
which can identify 
parts of a system that are most responsible for incorrect behaviour. 

In this paper, we introduce \textit{DeepFault}, the first fault 
localization-based whitebox testing approach for DNNs. The objectives of 
DeepFault are twofold: 
(i) \textit{identification} of \textit{suspicious} neurons, i.e., neurons 
likely to be more responsible for incorrect DNN behaviour; and 
(ii) \textit{synthesis} of new inputs, using correctly classified inputs, 
that exercise the identified suspicious neurons. 
Similar to conventional fault localization, which receives as input a faulty 
software and outputs a ranked list of suspicious code locations where the 
software may be defective~\cite{wong2016survey}, DeepFault \emph{analyzes} the 
behaviour of neurons of a DNN after training to establish their hit spectrum 
and \emph{identifies} suspicious neurons by employing suspiciousness measures. 
DeepFault employs a suspiciousness-guided 
algorithm to \emph{synthesize} new inputs, that achieve high activation values 
for suspicious neurons, by modifying correctly classified inputs. 
Our empirical evaluation on the popular publicly available datasets 
MNIST~\cite{lecun1998mnist} and CIFAR-10~\cite{cifar_model} provides evidence 
that DeepFault can identify neurons which can be held responsible for 
insufficient network performance. DeepFault can also synthesize new inputs, 
which closely resemble the original inputs, are highly adversarial and increase 
the activation values of the identified suspicious neurons.
To the best of our knowledge, DeepFault is the first research attempt that 
introduces \textit{fault localization} for DNNs to identify suspicious neurons 
and synthesize new, likely adversarial, inputs.

Overall, the main contributions of this paper are:
\vspace*{-1mm}
\begin{itemize}
	\item The DeepFault approach for whitebox testing of DNNs driven by fault 
	localization; 
	\item An algorithm for identifying suspicious neurons that 	adapts 
	suspiciousness measures from the domain of  spectrum-based fault 
	localization; 
	\item A suspiciousness-guided algorithm to synthesize inputs 
	that achieve high activation values of potentially suspicious neurons;
	\item A comprehensive evaluation of DeepFault on two public datasets (MNIST 
	and CIFAR-10) demonstrating its feasibility and effectiveness; 
	\vspace*{-1mm}
\end{itemize}

The reminder of the paper is structured as follows. 
Section~\ref{sec:background} presents briefly DNNs and fault 
localization in traditional software testing. Section~\ref{sec:approach} 
introduces \textit{DeepFault} and Section~\ref{sec:implementation} presents its 
open-source implementation. Section~\ref{sec:evaluation} describes the 
experimental setup, 
research questions and evaluation carried out. Sections~\ref{sec:relatedWork} 
and~\ref{sec:conclusion} discuss related work and conclude the paper, 
respectively.

\vspace*{-3mm}
\section{Background\label{sec:background}}
\vspace*{-2mm}

\subsection{Deep Neural Networks \label{ssec:dnns}}
\vspace*{-1mm}

\begin{wrapfigure}{R}{0.5\textwidth}
	\centering
	\vspace*{-20mm}
	\includegraphics[width=0.45\textwidth]{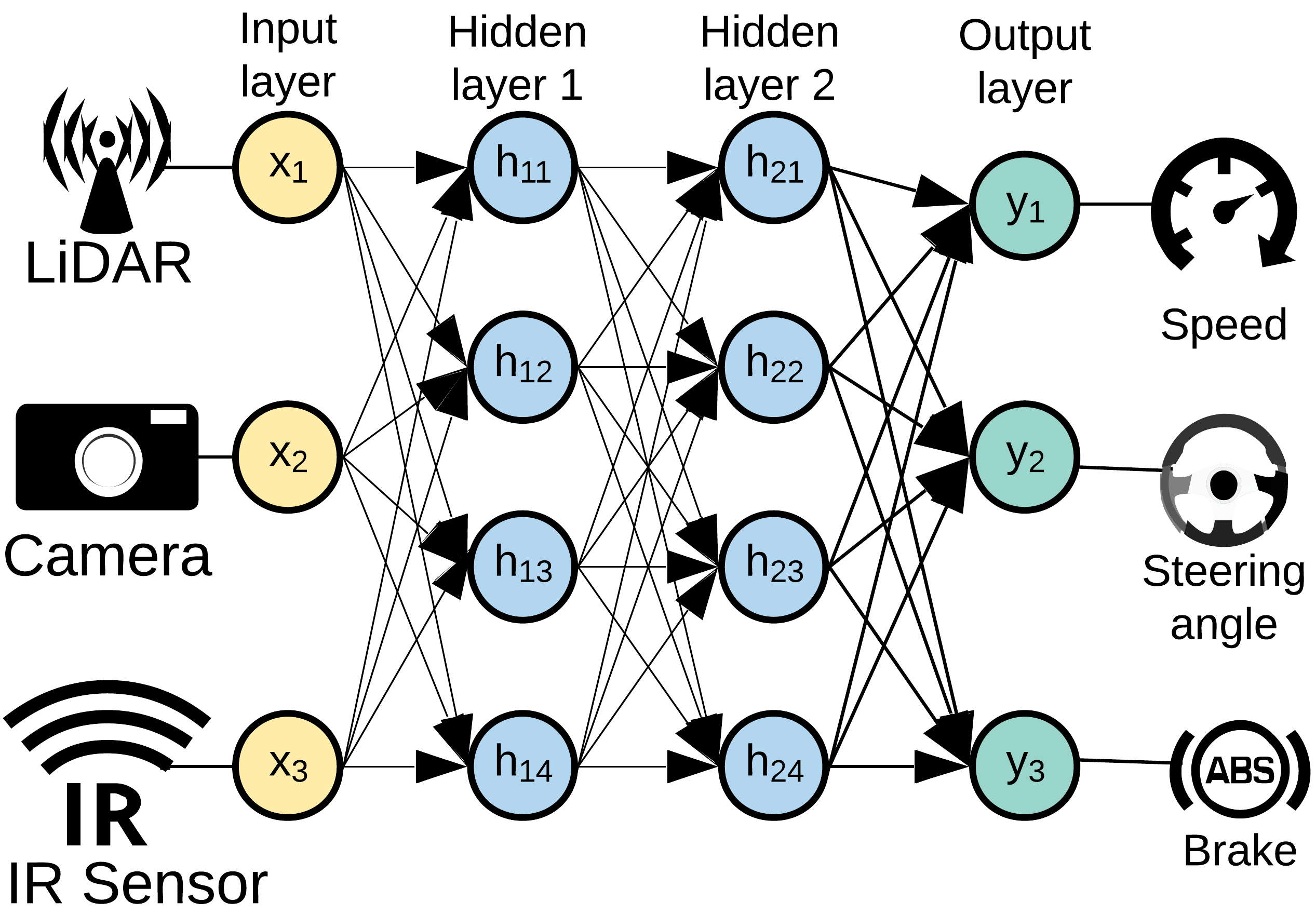}
	
	\vspace*{-2mm}
	\caption{A four layer fully-connected DNN that receives inputs from vehicle 
		sensors (camera, LiDAR, infrared) and outputs a decision for speed, 
		steering angle and brake.}
	\label{fig:nn}
	
	\vspace*{-6mm}
\end{wrapfigure}

We consider Deep Learning software systems in which one or more system 
modules is controlled by DNNs~\cite{Goodfellow-et-al-2016}. 
A typical feed-forward DNN comprises multiple interconnected 
neurons organised into several layers: the \emph{input} layer, the 
\emph{output} layer and at least one \emph{hidden} layer (Fig.~\ref{fig:nn}). 
Each DNN layer comprises a sequence of neurons. 
A \emph{neuron} denotes a computing unit that applies a \emph{nonlinear 
activation function} to its inputs and transmits the result to neurons in 
the successive layer. 
Commonly used activation functions are sigmoid, hyperbolic tangent, ReLU 
(Rectified Linear Unit) and leaky ReLU~\cite{Goodfellow-et-al-2016}. 
Except from the input layer, every neuron is connected to neurons in the 
successive layer with \emph{weights}, i.e., edges, 
whose values signify the strength of a connection between neuron pairs. 
Once the DNN architecture is defined, i.e., the number of layers, neurons per 
layer 
and activation functions, the DNN undergoes a \emph{training process} using a 
large amount of labelled training data to find weight values 
that minimise a \emph{cost function}. 

In general, a DNN could be considered as a parametric multidimensional function 
that consumes input data (e.g, raw image pixels) in its input layer,
extracts \textit{features}, i.e., semantic concepts, by performing a series of 
nonlinear transformations in its \emph{hidden layers}, and, finally,  produces 
a decision that matches the effect of these computations in its \emph{output 
layer}. 

\subsection{Software Fault Localization \label{ssec:faultLocalisation}}
\vspace*{-2mm}

\begin{figure}[h!]
		\vspace*{-4mm}
		
		\centering
		\includegraphics[width=\linewidth]{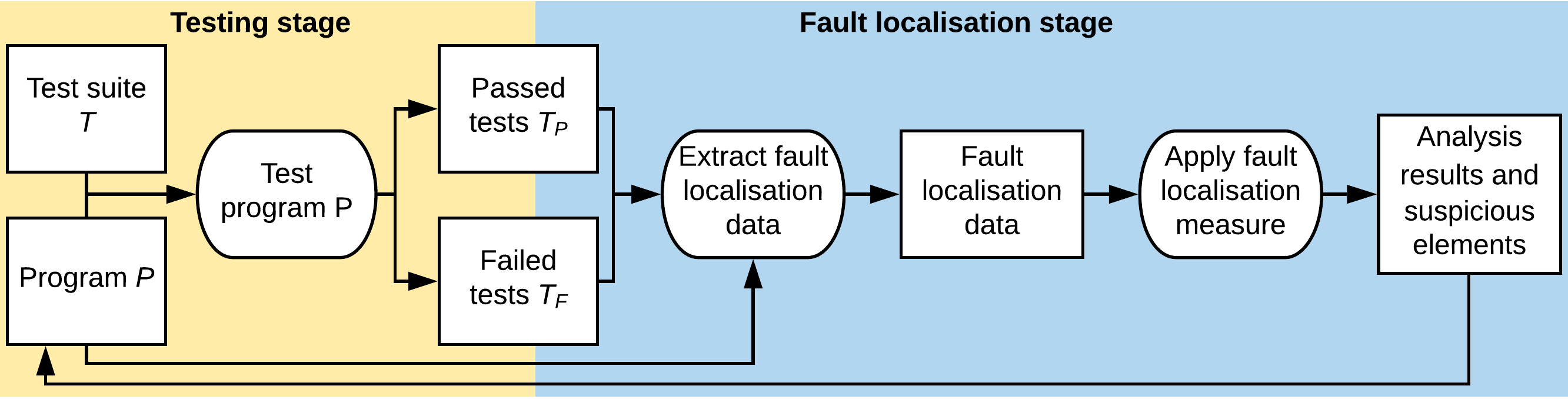}
		
		\vspace*{-2mm}
		\caption{General fault localization for traditional software systems.}
		\label{fig:traditionalFL}
		
	\end{figure}

Fault localization (FL) is a white box testing technique that focuses on
identifying source code elements 
(e.g., statements, declarations) that are more likely to contain faults.
The general FL process~\cite{wong2016survey} for traditional 
software (Fig.~\ref{fig:traditionalFL}) 
uses as inputs a program \emph{P}, corresponding to the system under test, and 
a test suite \emph{T}, and employs an FL technique to test \emph{P} against \emph{T} 
and establish subsets that represent the passed and failed tests. 
Using these sets and information regarding program elements $p \in P$, the FL 
technique extracts fault localization data  which is then employed by an 
FL measure to establish the ``suspiciousness'' of each program element $p$. 
Spectrum-based FL, the most studied class of FL techniques, uses program traces 
(called program spectra) of successful and failed test executions to establish 
for program element $p$ the tuple $(\!e_s,\!e_f,\!n_s,\!n_f\!)$.
Members $e_s$ and $e_f$ ($n_s$ and $n_f$) represent the number of times the 
corresponding program element has been (has not been) executed by tests, with 
success and fail, respectively. 
A spectrum-based FL measure consumes this list of tuples and ranks the program 
elements in decreasing order of suspiciousness enabling software engineers to 
inspect program elements and find faults effectively. 
For 
a comprehensive survey of state-of-the-art FL techniques, see~\cite{wong2016survey}.

\vspace*{-4mm}
\section{DeepFault\label{sec:approach}}

\begin{figure*}[t]
	\centering
	\includegraphics[width=\linewidth]{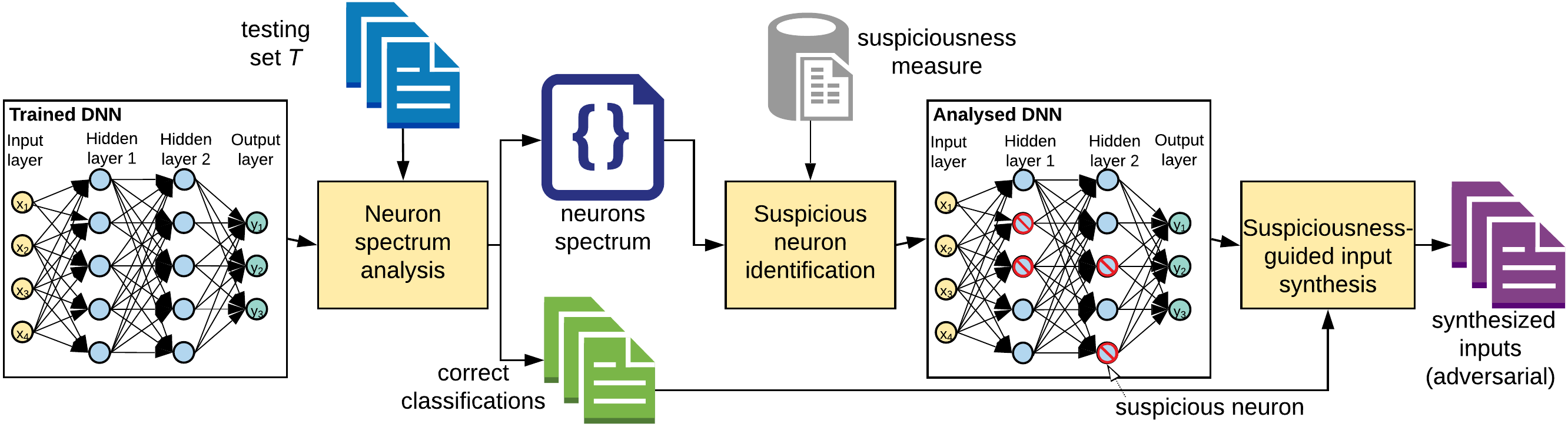}
	
	\vspace*{-2mm}
	\caption{DeepFault workflow.}
	
	\label{fig:overview}
	\vspace*{-4mm}
\end{figure*}

In this section, we introduce our DeepFault whitebox approach that enables to 
systematically test DNNs by identifying and localizing highly erroneous neurons 
across a DNN.
Given a pre-trained DNN, DeepFault, whose workflow is shown in 
Figure~\ref{fig:overview}, performs a series of \emph{analysis}, 
\emph{identification} and \emph{synthesis} steps to identify highly erroneous 
DNN neurons and synthesize new inputs that exercise erroneous neurons. 
We describe the DeepFault steps in 
Sections~\ref{ssec:analysis}--\ref{ssec:synthesis}.

We use the following notations to describe DeepFault. 
Let $\mathcal{N}$ be a DNN with $l$ layers. Each layer $L_i, 1\leq i \leq l$, 
consists of $s_i$ 
neurons and the total number of neurons in $\mathcal{N}$ is given by $s = 
\sum_{i=1}^l 
s_i$. Let also $n_{i,j}$ be the $j$-th neuron in the $i$-th layer. 
When the context is clear, we use $n \in \mathcal{N}$ to denote any neuron 
which is part of the DNN $\mathcal{N}$ irrespective of its layer. 
Likewise, we use $N_H$ to denote the neurons which belong to the hidden layers 
of N, i.e., $N_H = \{n_{ij} | 1 < i < l, 1\leq j \leq s_j\}$.
We use $\mathcal{T}$ to denote the set of test inputs from 
the input domain of $\mathcal{N}$, $t \in \mathcal{T}$ to denote a concrete 
input, and $u \in t$ for an element of $t$. 
Finally, we use the function $\phi(t, n)$ to signify the output of the 
activation function of neuron $n \in \mathcal{N}$.

\vspace*{-3mm}
\subsection{Neuron Spectrum Analysis\label{ssec:analysis}}
\vspace*{-2mm}
The first step of DeepFault involves the analysis of neurons within a DNN to 
establish suitable neuron-based attributes that will drive the detection and 
localization of faulty neurons. 
As highlighted in recent research~\cite{pei2017deepxplore,GJZCS18}, the 
adoption of whitebox testing techniques provides additional useful insights
regarding internal neuron activity and network behaviour. These insights 
cannot be easily extracted through black-box DNN testing, i.e., 
assessing the performance of a DNN considering only the decisions made given a 
set of test inputs $\mathcal{T}$.

DeepFault initiates the identification of suspicious neurons 
by establishing attributes that capture a neuron's execution pattern. 
These attributes are defined as follows.
Attributes $attr_n^{\text{as}}$ and $attr_n^{\text{af}}$ signify the 
number of 
times neuron $n$ was active (i.e., the result of the activation function 
$\phi(t, n)$ was above the predefined threshold) and the network made a 
successful or failed decision, respectively. 
Similarly, attributes $attr_n^{\text{ns}}$ and $attr_n^{\text{nf}}$ cover the 
case in which neuron $n$ is not active.
DeepFault analyses the behaviour of neurons in the DNN hidden layers, 
under a specific test set $\mathcal{T}\!\!$, to assemble a \emph{Hit Spectrum 
(HS)} 
for each neuron, i.e., a tuple describing its dynamic behaviour.
We define formally the HS 
as follows. 

\begin{definition}\label{def:spectrum}
	Given a DNN $\mathcal{N}$ and  a test set $\mathcal{T}$, 
	we say that for any neuron 
	$n \in \mathcal{N}_H$ 
	its hit spectrum is given by 
	the tuple $HS_n 	= (attr_n^\text{as}, 
	attr_n^\text{af},$ $attr_n^\text{ns}, attr_n^\text{nf})$. 
\end{definition} 

\noindent
Note that the sum of each neuron's HS should be equal to the size of 
$\mathcal{T}$.

Clearly, the interpretation of a hit spectrum (cf. Def.~\ref{def:spectrum}) is 
meaningful only for neurons in the hidden layers of a DNN. Since neurons within 
the input layer $L_1$ 
correspond to elements from the input domain (e.g., pixels from an image 
captured by a camera in Fig.~\ref{fig:nn}), we consider them to be 
``correct-by-construction''. 
Hence, these neurons cannot be credited or held responsible for a successful or 
failed decision made by the network. Furthermore, input neurons are always 
active and thus propagate one way or another their values to neurons in the 
following layer. 
Likewise, neurons within the output layer $L_l$ simply aggregate
values from neurons in the penultimate layer $L_{l-1}$, multiplied by the 
corresponding weights, and thus have limited influence in the overall network 
behaviour and, accordingly, to decision making. 


\vspace*{-3mm}
\subsection{Suspicious Neurons Identification\label{ssec:suspicious}}
\vspace*{-1mm}
During this step, DeepFault consumes the set of hit spectrums, derived from DNN 
analysis, and identifies \emph{suspicious} neurons which are likely to have 
made significant contributions in achieving inadequate DNN performance (low 
accuracy/high loss).
To achieve this identification, DeepFault employs a spectrum-based 
suspiciousness measure 
which computes a suspiciousness score per neuron using spectrum-related 
information. 
Neurons with the highest suspiciousness score are more likely to 
have been trained unsatisfactorily and, hence, contributing more to incorrect 
DNN decisions. This indicates that the weights of these neurons need further 
calibration~\cite{Goodfellow-et-al-2016}. 
We define neuron suspiciousness as follows.

\vspace*{-3mm}
\begin{definition}\label{def:susp}
	Given a neuron  $n \in \mathcal{N}_H$
	with $\textit{HS}_n$ being its hit spectrum, a neuron's 
	spectrum-based suspiciousness is given by the 
	function  $\textsc{Susp}_n \!:\! HS_n \!\rightarrow\!\! \mathbb{R}$.
	
	\vspace*{-4mm}
\end{definition}

Intuitively, a suspiciousness measure facilitates the derivation of 
correlations between a neuron's behaviour given a test set 
$\mathcal{T}$ and the failure pattern of $\mathcal{T}$ as determined by the 
overall network behaviour. 
Neurons whose behaviour pattern is \emph{close} to 
the failure pattern of $\mathcal{T}$ are more likely to operate unreliably,
and consequently, they should be assigned higher suspiciousness. 
Likewise, neurons whose behaviour pattern is \emph{dissimilar} to the failure 
pattern of $\mathcal{T}$ are considered more trustworthy and their 
suspiciousness values should be low.

In this paper, we instantiate DeepFault with three different suspiciousness 
measures, i.e., 
Tarantula~\cite{jones2005empirical}, Ochiai~\cite{ochiai1957zoogeographic} and 
D*~\cite{wong2014dstar} whose algebraic formulae are shown in 
Table~\ref{table:suspiciousnessMeasures}.
The general principle underlying these suspiciousness measures is that the more 
often a neuron is activated by test inputs for which the DNN made an incorrect 
decision, and the less often the neuron is activated by test inputs for which 
the DNN made a correct decision, the more suspicious the neuron is. 
These suspiciousness measures have been adapted from the domain of fault 
localization in software engineering~\cite{wong2016survey} in which 
they have achieved competitive results in automated software debugging by
isolating the root causes of software failures while reducing human input.
To the best of our knowledge, DeepFault is the first approach that 
proposes to incorporate these suspiciousness measures into the DNN domain for 
the identification of defective neurons.

\begin{table}[t]
	\caption{Suspiciousness measures used in DeepFault}
	\vspace*{-6mm}	

	\begin{center}
			\begin{tabular}{p{4.5cm} p{6cm}}
				\toprule
				\textbf{Suspiciousness Measure} & \textbf{Algebraic Formula}\\
				 Tarantula~\cite{jones2005empirical}: & 
				 	$\frac{attr_n^\text{af}/(attr_n^\text{af} + 
				 	attr_n^\text{nf})}{attr_n^\text{af}/(attr_n^\text{af} + 
				 	attr_n^\text{nf}) + attr_n^\text{as}/(attr_n^\text{as} + 
				 	attr_n^\text{ns})}$\\[2mm]
				 Ochiai~\cite{ochiai1957zoogeographic}: &
				 	$\frac{attr_n^\text{af}}
				 	{\sqrt{(attr_n^\text{af} + attr_n^\text{nf}) \cdot 
				 	 (attr_n^\text{af} + attr_n^\text{as})}}$\\[2mm]
				 D*~\cite{wong2014dstar}: & 
			 	 $\frac{attr_n^{\text{af}^*}}
			 	 {attr_n^\text{as}+attr_n^\text{nf}}$\\[1.1mm]
			\bottomrule
			\multicolumn{2}{l}{
			{\vspace*{-2em}\scriptsize $* > 0$ is a variable. We used $*=3$, 
			among the most widely explore 
			values~\cite{wong2016survey,pearson2017evaluating}}.}
		\end{tabular}
	
		\label{table:suspiciousnessMeasures}
		\vspace*{1mm}
	\end{center}
\end{table}

\setlength{\textfloatsep}{2mm}
\begin{algorithm}[t]
	\caption{Identification of suspicious neurons}\label{alg:identification}
	\begin{small}
		\renewcommand{\baselinestretch}{1}
		\begin{algorithmic}[1]
			\Function{SuspiciousNeuronsIdentification}{$\mathcal{N}, 
				\mathcal{T}, k$}
			\State $S \gets \emptyset$\label{a1:l1}
			\hfill\Comment suspiciousness vector
			\ForAll{$n\in\mathit{N}$}\label{a1:allN}
			\State $HS_n \gets \emptyset$\label{a1:l2}
			\hfill\Comment$n$-th neuron hit spectrum vector
			\ForAll{$p\in\{as, af, ns, nf\}$}\label{a1:allP}
			\State $a_n^p = $\textsc{attr}$(\mathcal{T},p)$
			\hfill\Comment establish attribute for property $p$
			\State $HS_n = HS_n \;\cup\; \{a_n^p\}$
			\hfill\Comment construct hit spectrum (cf. Def.~\ref{def:spectrum})
			\EndFor \label{a1:allPend}
			\State $S = S \;\cup\; \{\textsc{Susp}(\textit{HS}_n)\}$
			\hfill \Comment determine neuron suspiciousness (cf. 
			Def.~\ref{def:susp})
			\EndFor \label{a1:allNend}
			\State SN = $\{n | \textsc{Susp}(\textit{HS}_n) \in 
			\textsc{Select}(S, k)\}$\hfill \Comment 
			select the $k$ 
			most suspicious neurons
			\State \textbf{return} SN
			\EndFunction
		\end{algorithmic}
	\end{small}
\end{algorithm}

The use of suspiciousness measures in DNNs targets the identification of a set 
of defective neurons rather than diagnosing an isolated defective neuron.
Since the output of a DNN decision task is typically based on the 
aggregated effects of its neurons (computation units), with each neuron making 
its own contribution to the whole computation 
procedure~\cite{Goodfellow-et-al-2016}, identifying a single point of failure 
(i.e., a single defective neuron) has limited value.
Thus, after establishing the suspiciousness of neurons in the hidden layers 
of a DNN, the neurons are ordered in decreasing order of suspiciousness and the 
$k, 1 \leq l \leq s$,  most probably defective (i.e., ``undertrained'') 
neurons are selected. 
Algorithm~\ref{alg:identification} presents the high-level steps for 
identifying and selecting the $k$ most suspicious neurons.
When multiple neurons achieve the same suspiciousness score,
DeepFault resolves ties by prioritising neurons that belong to deeper hidden 
layers (i.e., they are closer to the output layer).
The rationale for this decision lies in fact that neurons in deeper layers are 
able to learn more meaningful representations of the input 
space~\cite{zeiler2014visualizing}.


\vspace*{-2mm}
\subsection{Suspiciousness-Guided Input Synthesis \label{ssec:synthesis}}
\vspace*{-2mm}

DeepFault uses the selected $k$ most suspicious neurons (cf. 
Section~\ref{ssec:suspicious}) to synthesize inputs that exercise 
these neurons and could be adversarial (see 
Section~\ref{sec:evaluation}). 
The premise underlying the synthesis is that increasing the activation 
values of suspicious neurons will cause the propagation of degenerate 
information, computed by these neurons, across the network, 
thus, shifting the decision boundaries in the output layer. 
To achieve this, DeepFault applies targeted modification of test inputs from 
the test set $\mathcal{T}$ for which the DNN made correct decisions (e.g., for 
a classification task, the DNN determined correctly their ground truth classes)
aiming to steer the DNN decision to a different
region (see Fig.~\ref{fig:overview}).

\setlength{\textfloatsep}{0mm}
\begin{algorithm}[t]
	\caption{New input synthesis guided by the identified suspicious neurons
	}\label{alg:syn}
	\begin{small}
		\renewcommand{\baselinestretch}{1}
		\begin{algorithmic}[1]
			\Statex \textbf{\hspace*{-1.1em}Input:} $SN \gets$ suspicious 
			neurons (Algorithm~\ref{alg:identification}), $step \gets$ step 
			size in gradient ascent
			\Statex \hspace*{-4mm}$T_s \!\gets\!$ test inputs correctly 
			classified by 	$\mathcal{N}$, 
			$d \!\gets\!$ new inputs maximum allowed distance
			\Function{SuspiciousnessGuidedInputSynthesis}{$SN, 
				\mathcal{T}_s, d, step$}
			\State $NT \gets \emptyset$\label{a2:l1} 
			\hfill\Comment set of synthesized inputs
			\ForAll{$t\in T_s$}\label{a2:1}
			\State $G_t \gets \emptyset$\label{a2:l2} 
			\hfill\Comment gradient collection of suspicious neurons
			\ForAll{$n\in SN$}\label{a2:l3}
			\State $n^{v} = \phi(t, n)$\label{al2:value}			
			\hfill\Comment determine output of neuron
			\State $G = \partial n^v/\partial t$\label{al2:gradient}
			\hfill\Comment establish gradient of neuron for $t$
			\State $G_t = G_t \cup \{G\}$ 
			\hfill\Comment collect gradients of suspicious neurons for $t$
			\EndFor
			\State $t' \gets \emptyset$ 
			\hfill\Comment initialisation of input to be synthesised
			\ForAll{$u\in t$}\label{a2:l4}
			\State $u_{gradient} = \sum_{G \in G_t} G / |G_t|$
			\hfill\Comment determine average gradient of 
			$u$\label{al2:avgGrad}
			\State $u_{gradient} = 
			\textsc{GradientConstraint}(u_{gradient},d, step)$ 
			\label{al2:ugrad}
			\State $t' = t' \frown \{\textsc{DomainConstraints}(u + 
			u_{gradient})\}$\label{al2:newt}
			\EndFor
			\State $NT = NT \cup \{t'\}$
			\EndFor
			\State \textbf{return} $NT$
			\EndFunction
		\end{algorithmic}
	\end{small}
\end{algorithm}

Algorithm~\ref{alg:syn} shows the high-level process for synthesising new 
inputs based on the identified suspicious neurons. 
The synthesis task is underpinned by a gradient ascent algorithm that 
aims at determining the extent to which a correctly classified 
input should be modified to increase the activation values of suspicious 
neurons. 
For any test input $t \in T_s$ correctly classified by the DNN, we extract 
the value of each suspicious neuron and its gradient in lines~\ref{al2:value} 
and~\ref{al2:gradient}, respectively. Then, by iterating over each input 
dimension $u \in t$, we determine the gradient value $u_{gradient}$ by which 
$u$ will be perturbed (lines~\ref{al2:avgGrad}-\ref{al2:ugrad}). 
The value of $u_{gradient}$ is based on the mean gradient of $u$ across the 
suspicious neurons controlled by the function \textsc{GradientConstraints}.
This function uses a test set specific $step$ parameter and a distance $d$ 
parameter to facilitate the synthesis of realistic test inputs that are 
sufficiently \emph{close}, according to $L_\infty$-norm, to the original 
inputs. 
We demonstrate later in the evaluation of DeepFault (cf. 
Table~\ref{table:similarity})  that these parameters enable the synthesis of 
inputs similar to the original.
The function \textsc{DomainConstraints} applies domain-specific constraints 
thus ensuring that $u$ changes due to gradient ascent result in realistic and 
physically reproducible test inputs as in~\cite{pei2017deepxplore}. 
For instance, a domain-specific constraint for an image classification dataset
involves bounding the pixel values of synthesized images to be within a certain 
range (e.g., 0--1 for the MNIST dataset~\cite{lecun1998mnist}).
Finally, we append the updated $u$ to construct a new test input $t'$ 
(line~\ref{al2:newt}).

As we experimentally show in Section~\ref{sec:evaluation}, the 
suspiciousness measures used by DeepFault can synthesize adversarial inputs 
that cause the DNN to misclassify previously correctly 
classified inputs.
Thus, the identified suspicious neurons can be attributed a degree of 
responsibility for the inadequate network performance meaning that their 
weights have not been optimised. This reduces the DNN's ability
for high generalisability and correct 
operation in untrained data. 

\vspace*{-3mm}
\section{Implementation\label{sec:implementation}}
\vspace*{-4mm}
To ease the evaluation and adoption of the DeepFault approach (cf. 
Fig.~\ref{fig:overview}), we have implemented 
a prototype tool on top of the open-source machine learning framework Keras 
(v2.2.2)~\cite{chollet2015keras} with Tensorflow(v1.10.1) backend~\cite{tensorflow}. 
The full experimental results are summarised in the following section. 

\vspace*{-2mm}
\section{Evaluation\label{sec:evaluation}}
\vspace*{-2mm}

\subsection{Experimental Setup}
\vspace*{-2mm}
We evaluate DeepFault on two popular publicly available datasets. 
MNIST~\cite{lecun1998mnist} is a  handwritten digit dataset with 60,000 
training samples 
and 10,000 testing 
samples; each input is a 28x28 pixel image with a class label from 0 to 9.  
CIFAR-10~\cite{cifar_model}) is an image dataset with 50,000 training samples 
and 10,000 
testing samples; each input is a 32x32 image in ten different classes (e.g., 
dog,$\!$ 
bird,$\!$ car).

For each dataset, we study three DNNs that have been used in previous 
research~\cite{wicker2018feature,cifar_model} (Table~\ref{table:dnns}).
All DNNs have different architecture and number of trainable parameters.
For MNIST, we use fully connected neural networks (dense) and for CIFAR-10 we 
use 
convolutional neural networks with max-pooling and dropout layers that have 
been trained to 
achieve at least 95\% and 70\% accuracy on the provided test sets, 
respectively.  
The column `Architecture' shows the number of fully connected hidden layers and 
the 
number of neurons per layer. 
Each DNN uses a leaky ReLU \cite{maas2013rectifier} as its activation function
$(\alpha\!=\!0.01)$, which has been shown to achieve competitive accuracy 
results~\cite{xu2015empirical}. 

We instantiate DeepFault using the suspiciousness measures 
Tarantula~\cite{jones2005empirical}, Ochiai~\cite{ochiai1957zoogeographic} and 
D*~\cite{wong2014dstar} (Table~\ref{table:suspiciousnessMeasures}).
We analyse the effectiveness of DeepFault instances using different number of 
suspicious 
neurons, i.e., $k\!\in\! 
\{1,2,3, 5, 10\}$ and $k \!\in\! \{10, 20, 30, 40, 50\}$ for MNIST and CIFAR 
models, 
respectively. We also ran preliminary experiments for each model from 
Table~\ref{table:dnns} to tune the hyper-parameters of Algorithm~\ref{alg:syn} 
and 
facilitate replication of our findings. 
Since gradient values are model and input specific, the perturbation magnitude 
should reflect these values and reinforce their impact. 
We determined empirically that $step\!=\!1$ and $step\!=\!10$ are good values, 
for MNIST 
and CIFAR models, respectively,  that enable our algorithm to perturb inputs. 
We also set the maximum allowed distance $d$ to be at most $10\%$ ($L_\infty$) 
with regards 
to the range of each input dimension (maximum pixel value).
As shown in Table~\ref{table:similarity}, the synthesized inputs are very 
similar to the 
original inputs and are rarely constrained by $d$.
Studying other $step$ and $d$ values is part of our future work. 
All experiments were run on an Ubuntu server with 16 GB memory and Intel Xeon E5-2698 2.20GHz.\\

\begin{table}[t]
	\fontsize{9}{11}\selectfont
	\caption{Details of MNIST and CIFAR-10 DNNs  used in the evaluation.}
	\vspace*{-6mm}
	\begin{center}\renewcommand{\arraystretch}{0.8}
		\begin{tabular}{ccccc}
			\toprule
			\textbf{Dataset} & \textbf{Model Name} & \textbf{\# Trainable 
			Params} & \textbf{Architecture} & \textbf{Accuracy} \\ 
			\midrule
			\multirow{3}{*}{MNIST} & MNIST\_1 & 27,420 & \textless$5 \times 
			30$\textgreater &  96.6\% \\ 
			& MNIST\_2 & 22,975 & \textless$6 \times 25$\textgreater  & 95.8\% 
			\\ 
			& MNIST\_3 & 18,680 & \textless$8 \times 20$\textgreater &  95\%\\ 
			\midrule
			\multirow{3}{*}{CIFAR-10}  
			& CIFAR\_1 & 411,434 & \textless$4 \times 128$\textgreater & 
			70.1\%\\ 			
			& CIFAR\_2 & 724,010 & \textless$2 \times 256$\textgreater & 72.6\% 
			\\
			& CIFAR\_3 & 1,250,858 & \textless $1 \times 512$\textgreater & 
			76.1\% \\   
			\bottomrule
		\end{tabular}
	
		\vspace*{-2mm}
		\label{table:dnns}
	\end{center}
\end{table}

\vspace*{-2em}
\subsection{Research Questions \label{ssec:researchQuestions}}
\vspace*{-4mm}
Our experimental evaluation aims to answer the following research questions.
\vspace*{-5mm}
\squishlist
	\item [\textbf{RQ1 (Validation)}:]
	\textbf{Can DeepFault find suspicious neurons effectively?}
	If suspicious neurons do exist, suspiciousness measures used by DeepFault 
	should comfortably outperform a random suspiciousness selection strategy.

	\item [\textbf{RQ2 (Comparison)}:]
	\textbf{How do DeepFault instances using different suspiciousness measures 
	compare 
	against each other?}
	Since DeepFault can work with multiple suspiciousness measures, we examined 
	the results produced by DeepFault instances using 
	Tarantula~\cite{jones2005empirical}, 
	Ochiai~\cite{ochiai1957zoogeographic} and D*~\cite{wong2014dstar}.
	
	\item [\textbf{RQ3 (Suspiciousness Distribution)}:]
	\textbf{How are suspicious neurons found by DeepFault distributed across a 
	DNN?}
	With this research question, we analyse the distribution of suspicious 
	neurons in 
	hidden DNN layers using different suspiciousness measures.
		
	\item [\textbf{RQ4 (Similarity)}:]
	\textbf{How realistic are inputs synthesized by DeepFault?}
 	We analysed the distance between synthesized and original inputs to examine 
 	the extent to which DeepFault synthesizes realistic inputs.
	 	
	\item [\textbf{RQ5 (Increased Activations)}:]
	\textbf{Do synthesized inputs increase activation values of suspicious 
	neurons?}
	We assess whether the suspiciousness-guided input synthesis algorithm 
	produces inputs 
	that reinforce the influence of suspicious neurons across a DNN.
	
	\item [\textbf{RQ6 (Performance)}:]
	\textbf{How efficiently can DeepFault synthesize new inputs?}
	We analysed the time consumed by DeepFault to synthesize new inputs and the 
	effect of suspiciousness measures used in DeepFault instances.
\squishend

\subsection{Results and Discussion}
\vspace*{-2mm}
\subsubsection{\textbf{RQ1 (Validation)}.}
We apply the DeepFault workflow to the DNNs from Table~\ref{table:dnns}. 
To this end, we instantiate DeepFault with a suspiciousness measure, 
\emph{analyse} a 
pre-trained DNN given the dataset's test set $\mathcal{T}$, \emph{identify} 
$k$ neurons 
with the highest suspiciousness scores and 
\emph{synthesize} new inputs, from \emph{correctly classified} inputs, that 
exercise these 
suspicious neurons. 
Then, we measure the prediction performance of the DNN on the synthesized 
inputs using  the 
standard performance metrics: 
cross-entropy \emph{loss}, i.e., the divergence between output and target 
distribution, and 
\emph{accuracy}, i.e., the percentage of correctly classified inputs over all 
given inputs. 
Note that DNN analysis is done per class, since the activation pattern of 
inputs from the same class is similar to each 
other~\cite{zeiler2014visualizing}. 

Table~\ref{table:models_validation_avg} shows the average loss and accuracy for 
inputs 
synthesized by DeepFault instances using Tarantula (T), Ochiai (O), DStar (D) 
and a random 
selection strategy (R) for different number of suspicious neurons $k$ on the 
MNIST (top) 
and CIFAR-10 (bottom) models from Table~\ref{table:dnns}. 
Each cell value in Table~\ref{table:models_validation_avg}, except from random 
R, is 
averaged over 100 synthesized inputs (10 per class).
For R\hspace*{-0.1mm},\hspace*{-0.3mm} we collected 500 synthesized inputs 
(50 per class) 
over five independent runs, thus, reducing the risk that our findings may have 
been 
obtained by chance. 

As expected (see Table~\ref{table:models_validation_avg}), DeepFault using any 
suspiciousness measure (T, O, D) obtained considerably lower 
prediction performance than R on MNIST models. 
The suspiciousness measures T and O are also effective on CIFAR-10 model, 
whereas the performance between D and R is similar. 
These results show that the identified $k$ neurons are actually 
\textit{suspicious} and, hence, their weights are insufficiently trained. 
Also, we have sufficient evidence that increasing the activation value of 
suspicious neurons by slightly perturbing inputs that have been classified 
correctly by the DNN could transform them into adversarial.

We applied the non-parametric statistical test Mann-Whitney with 95\% 
confidence level~\cite{wohlin2012experimentation} to check for 
statistically significant performance difference between the various DeepFault 
instances and random. 
We confirmed the significant difference among T-R and O-R (p-value$\!<\!0.05$) 
for all MNIST and CIFAR-10 models and for all $k$ values. We also confirmed the 
interesting observation that significant difference between D-R exists only for 
MNIST models (all $k$ values). 
We plan to investigate this observation further in our future work.

Another interesting observation from Table~\ref{table:models_validation_avg} is 
the small performance difference of DeepFault instances for 
different $k$ values. 
We investigated this further by analyzing the activation values of the next 
$k'$ most suspicious neurons according to the suspiciousness order given by
Algorithm~\ref{alg:identification}. 
For instance, if $k\!=\!2$ we analysed the activation values of the next 
$k'\in\{3,,5,10\}$ most suspicious neurons. 
We observed that the synthesized inputs frequently increase the activation 
values of the $k'$ neurons whose suspiciousness scores are also high, in 
addition to increasing the values of the top $k$ suspicious neurons.

Considering these results, we have empirical evidence about the existence of 
\textit{suspicious} neurons which can be responsible for 
inadequate DNN performance. 
Also, we confirmed that DeepFault instances using sophisticated suspiciousness 
measures significantly outperform a random strategy for most of the 
studied DNN models (except from the D-R case on CIFAR models; see RQ3).

\begin{table*}[t]
	\centering
	\caption{Accuracy and loss of inputs synthesized by DeepFault on MNIST 
		(top) and 
		CIFAR-10 (bottom) datasets.
		The best results per suspiciousness measure are shown in bold.
		($k\!\!:\!\!\#$suspicious neurons, T:Tarantula, O:Ochiai, D:D*, 
		R:Random)}
	\vspace*{-3mm}
	\scalebox{0.9}{
		\begin{tabular}{|c|c|c|c|c|c?c|c|c|c?c|c|c|c|}%
			\hline%
			\textbf{$k$} & \textbf{Measure} & 
			\multicolumn{4}{|c|}{\textbf{MNIST\_1}}&\multicolumn{4}{|c|}{\textbf{MNIST\_2}}&\multicolumn{4}{|c|}{\textbf{MNIST\_3}}\\%
			\hline%
			&&T&O&D&R&T&O&D&R&T&O&D&R\\%
			\hline%
			\multirow{2}{*}{\textbf{1}} & Loss & 3.55 & 
			\textbf{6.19} & 4.03 & 2.42 & 3.48 & 
			3.53 & \textbf{3.97} & 2.78 &  7.35 & 
			\textbf{8.23} & 6.36 & 3.66
			\\ 
			& Accuracy & 0.26 & \textbf{0.16} & 0.2 
			& 0.59 & 0.3 & \textbf{0.2} & 0.5 & 0.49 
			& 0.16 & \textbf{0.1} & 0.13 & 0.39 \\ 
			\hline
			
			\multirow{2}{*}{\textbf{2}} & Loss & 3.73 & 
			\textbf{6.08} & 3.18 & 2.67 &  3.12 & 
			3.76 & \textbf{4.08} & 0.9 &  4.27 & 
			\textbf{6.81} & 6.5 & 3.06 \\ 
			& Accuracy & \textbf{0.16} & 0.23 & 0.4 
			& 0.58 &  0.23 &0.23 & \textbf{0.13} & 
			0.77 &  0.29 & \textbf{0.13} & 0.26 & 
			0.56\\ \hline
			
			\multirow{2}{*}{\textbf{3}} & Loss & 4.1 & 6.19 & 
			\textbf{6.25} & 1.14 & 2.39 & 
			\textbf{3.94} & 3.04 & 1.61 &  3.33 & 
			\textbf{7.59} & 6.98 & 2.91 \\
			& Accuracy & \textbf{0.23} & 
			\textbf{0.23} & 0.33 & 0.77 &  0.46 
			&0.26 & \textbf{0.23} & 0.67 & 0.26 & 
			\textbf{0.06} & 0.16 & 0.61\\ \hline
			
			\multirow{2}{*}{\textbf{5}} & Loss & 4.63 & 6.68 & 
			\textbf{6.97} & 1.1 &  2.49 & 
			\textbf{3.64} & 3.48 & 0.94 &  4.15 & 
			\textbf{7.22} & 6.47 & 1.22 \\ 
			& Accuracy & 0.23 & 0.23 & \textbf{0.13} 
			& 0.79 & 0.26 & 0.26 & \textbf{0.2} & 
			0.73 & 0.16 & \textbf{0.1} & 0.26 & 0.77 
			\\ \hline
			
			\multirow{2}{*}{\textbf{10}} & Loss & 4.97 & 6.95 & 
			\textbf{7.4} & 1.3 &  2.08 & 3.06 & 
			\textbf{3.82} & 0.49 &  4.45 & 
			\textbf{7.16} & 5.9 & 0.57 \\ 
			& Accuracy& 0.23  & \textbf{0.2} & 0.23 
			& 0.75 & 0.4 & \textbf{0.23} &0.26 & 
			0.86 & \textbf{0.13} & 
			\textbf{0.13} & 
			\textbf{0.13} & 0.87 \\ \hline
	\end{tabular}}

	\vspace*{1mm}
	
	\scalebox{0.9}{
		\begin{tabular}{|c|c|c|c|c|c?c|c|c|c?c|c|c|c|}%
			\hline%
			\textbf{$k$} & \textbf{Measure} & 
			\multicolumn{4}{|c|}{\textbf{CIFAR\_1}}&\multicolumn{4}{|c|}{\textbf{CIFAR\_2}}&\multicolumn{4}{|c|}{\textbf{CIFAR\_3}}\\%
			\hline%
			&&T&O&D&R&T&O&D&R&T&O&D&R\\%
			\hline%
			\multirow{2}{*}{\textbf{10}} & Loss & 12.75 & 
			\textbf{13.49} & 1.33 & 3.25 & 
			\textbf{8.42} & 8.41 & 0 & 2.49 & 
			\textbf{6.12} & 1.77 & 1.12 & 1.21
			\\ 
			& Accuracy & 0.2 & \textbf{0.16} & 0.9 & 
			0.79 & \textbf{0.47} & 
			\textbf{0.47} & 1.0 & 0.84 &  
			\textbf{0.62} & 0.88 & 0.92 & 0.91 \\ 
			\hline
			
			\multirow{2}{*}{\textbf{20}} & Loss & 
			\textbf{12.79} & 12.43 & 0.45 & 1.8 & 
			\textbf{8.81} & 6.92 & 0.32 & 1.67 & 
			\textbf{6.12} & 1.12 & 0.96 & 0.64 \\ 
			& Accuracy & \textbf{0.2} & 0.22 & 0.96 
			& 0.88 & \textbf{0.44} & 0.55 & 0.97 & 
			0.89 & \textbf{0.62} & 0.92 & 0.93 & 
			0.95\\ \hline
			
			\multirow{2}{*}{\textbf{30}} & Loss & 
			\textbf{13.19} & 13.13 & 0.38 & 1.43 & 
			\textbf{8.35} & 6.32 & 0.55 & 0.86 & 
			\textbf{5.64} & 0.76 & 0.42 & 0.41 \\
			& Accuracy & \textbf{0.18} & 
			\textbf{0.18} & 0.95 & 0.9 & 
			\textbf{0.48} & 0.6 & 0.95 & 0.94 & 
			\textbf{0.64} & 0.93 & 0.96 & 0.97\\ 
			\hline
			
			\multirow{2}{*}{\textbf{40}} & Loss & 
			\textbf{13.69} & 11.92 & 0.8 & 1.29 & 
			\textbf{9.4} & 5.01 & 0.32 & 0.61 & 
			\textbf{4.51} & 1.12 & 0.22 & 0.54 \\ 
			& Accuracy & \textbf{0.14} & 0.26 & 0.92 
			& 0.91 & \textbf{0.41} & 0.68 & 0.97 & 
			0.95 & \textbf{0.72} & 0.92 & 0.97 & 
			0.96 \\ \hline
			
			\multirow{2}{*}{\textbf{50}} & Loss & 12.1 & 
			\textbf{13.37} & 0.36 & 0.9 & 
			\textbf{9.59} & 3.38 & 0 & 0.56 & 
			\textbf{4.67} & 0.04 & 0.64 & 0.48 \\ 
			& Accuracy & 0.24 & \textbf{0.17} & 0.96 
			& 0.94 & \textbf{0.4} & 0.78 & 1.0 & 
			0.96 & \textbf{0.71} & 0.98 & 0.96 & 
			0.96 \\ \hline
	\end{tabular}}
	
	\vspace*{2mm}
	\label{table:models_validation_avg}%
\end{table*}

\subsubsection{\textbf{RQ2 (Comparison).}}
We compare DeepFault instances using different suspiciousness measures and 
carried out pairwise comparisons using the Mann-Whitney test to check for 
significant difference between T, O, and D$^*$. 
We show the results of these comparisons in Appendix.
We observe that Ochiai achieves better results on MNIST\_1 and MNIST\_3 models 
for various $k$ values. 
This result suggests that the suspicious neurons reported by Ochiai are more 
responsible for insufficient DNN performance.
D$^*$ performs competitively on MNIST\_1 and MNIST\_3 for $k \in \{3,5,10\}$, 
but its performance on CIFAR-10 models is significantly inferior to Tarantula 
and Ochiai. 
The best performing suspiciousness measure in CIFAR models for most $k$ values 
is, by a great amount, Tarantula.

These findings show that multiple suspiciousness measures could be used for 
instantiating DeepFault with competitive performance. 
We also have evidence that DeepFault using D$^*$ is ineffective for some 
complex networks (e.g., CIFAR-10), but there is insufficient evidence 
for the best performing DeepFault instance. 
Our findings conform to the latest research on software fault 
localization which claims that there is no single best spectrum-based 
suspiciousness 
measure~\cite{pearson2017evaluating}.

\vspace*{-4mm}
\subsubsection{\textbf{RQ3 (Suspiciousness Distribution).}} 
\begin{figure*}[t]
	\subfloat{\label{rev}
		\includegraphics[trim={2mm 0cm 0cm 30mm}, clip, 
		width=.51\textwidth]{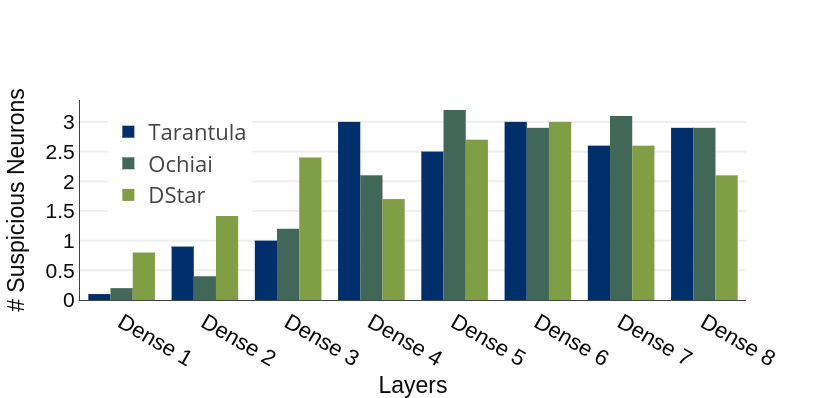}}
	\hspace*{-6mm}
	\subfloat
	{\label{rev_sol}
		\includegraphics[trim={10mm 0cm 8mm 32mm}, clip, 
		width=.51\textwidth]{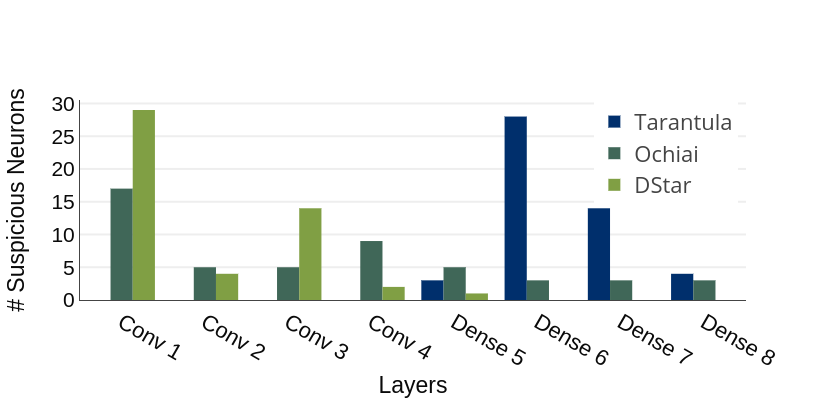}}
	
	\vspace*{-4mm}
	\caption{Suspicious neurons distribution on MNIST\_3  (left) and 
		CIFAR\_3 (right) models.}
	
	\label{fig:dist_susp}
	\vspace*{1mm}
\end{figure*}

We analysed the distribution of suspicious neurons identified by DeepFault 
instances across the hidden DNN layers.
Fig.~\ref{fig:dist_susp} shows the distribution of suspicious neurons on 
MNIST\_3 and CIFAR\_3 models with $k\!=\!10$ and $k\!=\!50$, respectively. 
Considering MNIST\_3, the majority of suspicious neurons are located at the 
deeper hidden layers (Dense 4-Dense 8) irrespective of the suspiciousness 
measure used by DeepFault. 
This observation holds for the other MNIST models and $k$ values. 
On CIFAR\_3, however, we can clearly see variation in the distributions 
across the suspiciousness measures. 
In fact, D$^*$ suggests that most of the suspicious neurons belong to initial 
hidden layers which is in contrast with Tarantula's recommendations. 
As reported in RQ2, the inputs synthesized by DeepFault using Tarantula 
achieved the best results on CIFAR models, thus showing that the identified 
neurons are actually suspicious. This difference in the distribution of 
suspicious neurons explains the inferior inputs synthesized by D$^*$ on CIFAR 
models (Table~\ref{table:models_validation_avg}). 

Another interesting finding concerns the relation between the suspicious 
neurons distribution and the ``adversarialness'' of synthesized inputs. 
When suspicious neurons belong to deeper hidden layers, the likelihood of the 
synthesized input being adversarial increases (cf. 
Table~\ref{table:models_validation_avg} and Fig.~\ref{fig:dist_susp}). 
This finding is explained by the fact that initial hidden layers transform 
input features (e.g., pixel values) into abstract features, while deeper 
hidden layers extract more semantically meaningful features and, thus, 
have higher influence in the final decision~\cite{Goodfellow-et-al-2016}. 

\vspace*{-6mm}
\subsubsection{\textbf{RQ4 (Similarity).}}
\begin{figure}[t]
	\centering
	\vspace*{-4mm}
	\includegraphics[scale=0.23, clip, trim=10mm 60mm 10mm 8mm] 
	{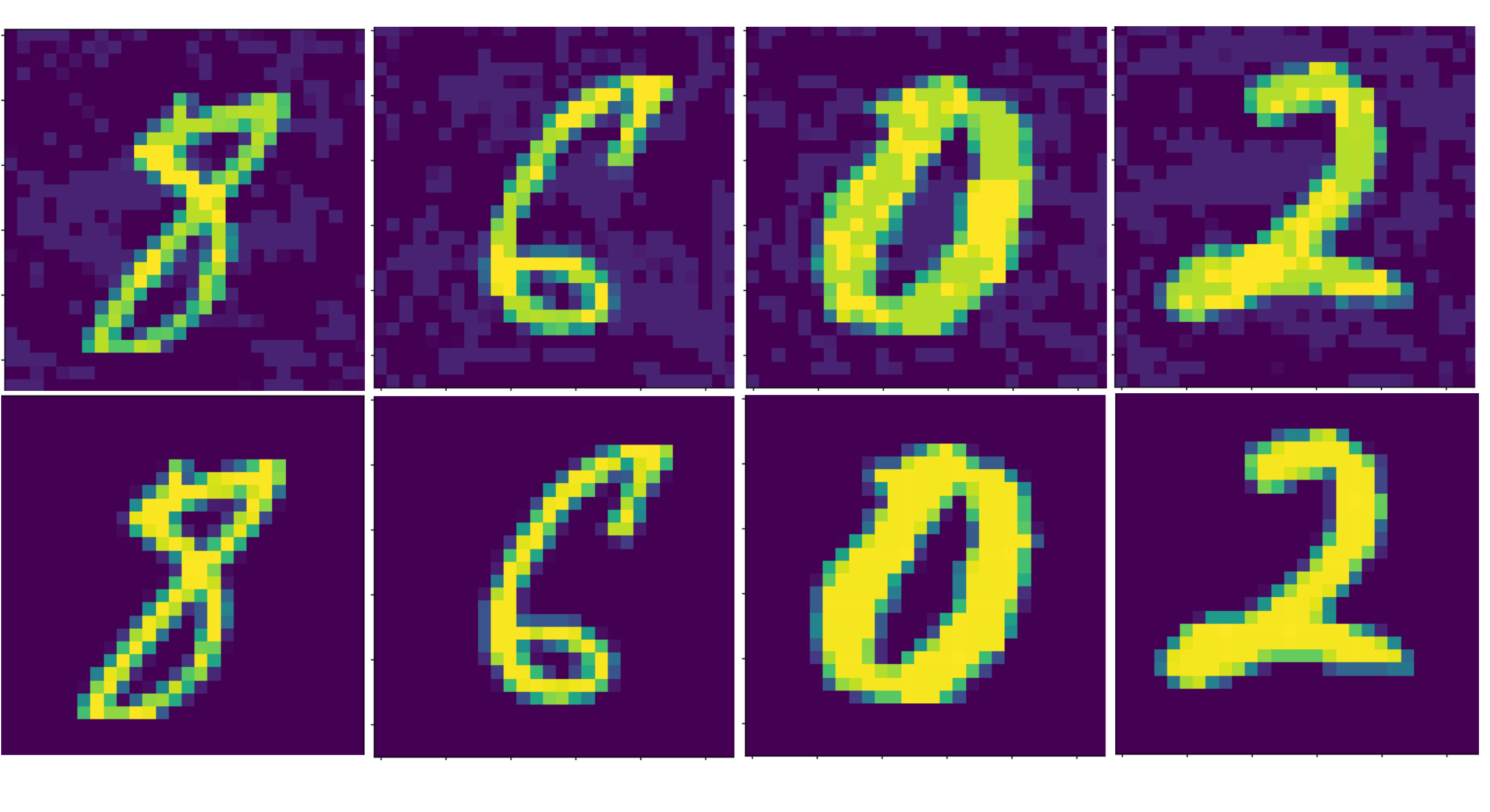}
	\includegraphics[scale=0.22, clip, trim=1mm 65mm 1mm 0mm] 	
	{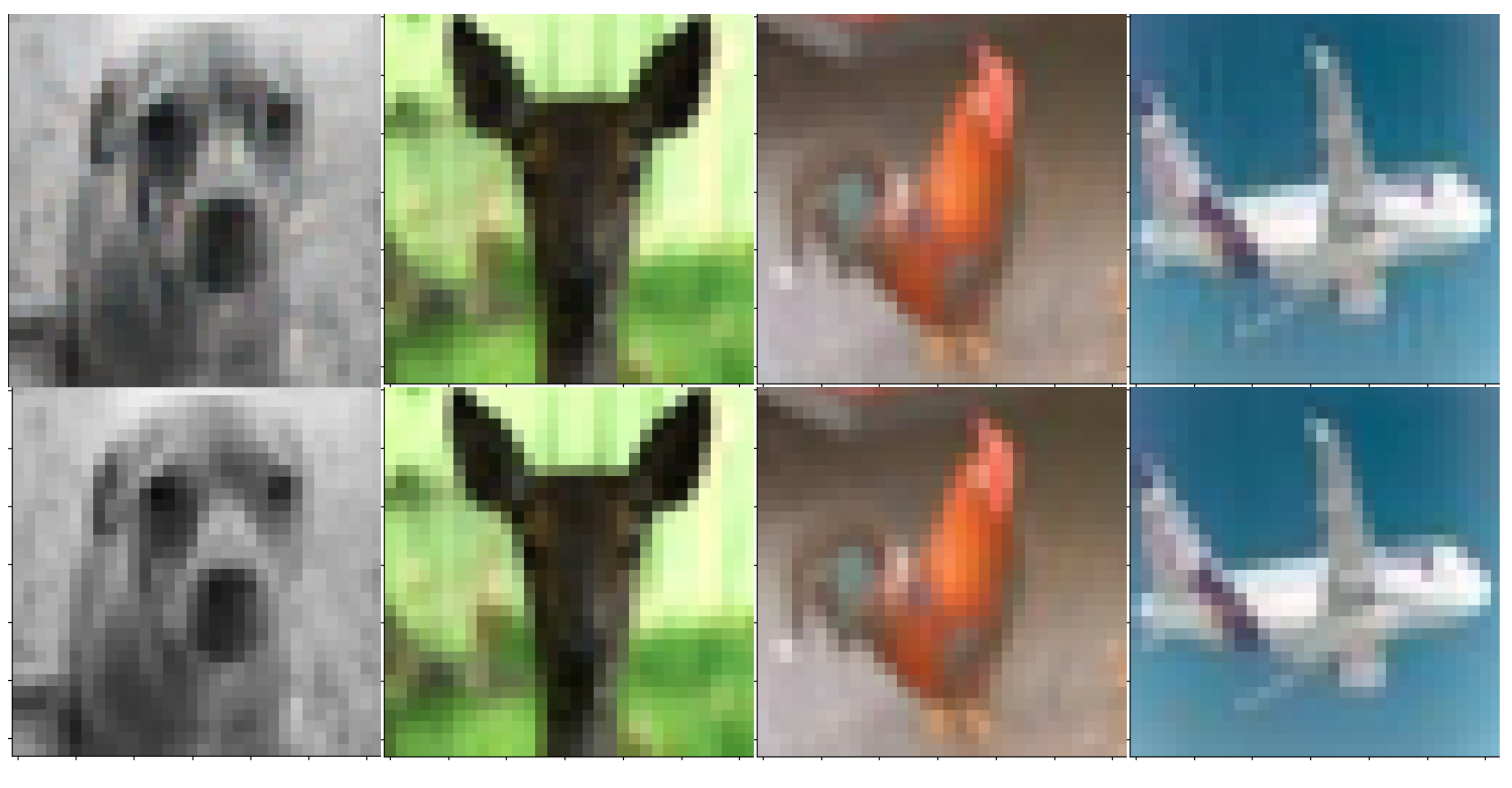}
	
	\vspace*{-5mm}
	\caption{\hspace*{-.9mm}Synthesized\hspace*{-.5mm} images\hspace*{-.5mm} 
		(top)\hspace*{-.5mm} 
		and\hspace*{-.5mm} 
		their\hspace*{-.5mm} originals\hspace*{-.5mm} (bottom).\hspace*{-.5mm}
		For\hspace*{-.5mm} each\hspace*{-.5mm} dataset\hspace*{-.1mm},
		suspi- cious\hspace*{-.3mm} neurons\hspace*{-.3mm} 
		are\hspace*{-.3mm} found\hspace*{-.3mm} 
		using\hspace*{-.3mm} (from\hspace*{-.3mm} left\hspace*{-.3mm} 
		to\hspace*{-.3mm} 
		right\hspace*{-.4mm}) Tarantula\hspace*{-.4mm}, Ochiai\hspace*{-.4mm}, 
		Dstar\hspace*{-.4mm} and\hspace*{-.4mm} Random.}
	\label{fig:pert_img}
	\vspace*{1.5mm}
\end{figure}

We examined the distance between original, correctly classified, inputs and 
those synthesized by DeepFault, to establish DeepFault's ability to synthesize 
realistic inputs. 
Table~\ref{table:similarity} (left) shows the distance between original and 
synthesized inputs for various distance metrics ($L_1$ Manhattan, $L_2$ 
Euclidean, $L\infty$ Chebyshev) for different $k$ values (\# suspicious 
neurons). The distance values, averaged over inputs synthesized using the 
DeepFault suspiciousness measures (T, O and D$^*$), demonstrate that the degree 
of perturbation  is similar irrespective of $k$ for MNIST models, whereas for 
CIFAR models the distance decreases as $k$ increases. 
Given that a MNIST input consists of 784 pixels, with each pixel taking values 
in $[0,1]$, the average perturbation per input is less than $5.28\%$ of the 
total possible perturbation ($L_1$ distance). 
Similarly, for a CIFAR input that comprises 3072 pixels, with each pixel taking 
values in $\{0,\!1,\!...,\!255\}$, the average perturbation per input is less 
that $0.03\%$ of the total possible perturbation ($L_1$ distance).  
Thus, for both datasets, the difference of synthesized inputs to their original 
versions is very small. 
We qualitatively support our findings by showing in Fig.~\ref{fig:pert_img} the 
synthesized images and their originals for an example set of inputs from the 
MNIST and CIFAR-10 datasets. 

We also compare the distances between original and synthesized inputs based on 
the suspiciousness measures (Table~\ref{table:similarity} right). 
The inputs synthesized by DeepFault instances using T, O or D$^*$ are very 
close to the inputs of the random selection strategy ($L_1$ 
distance). 
Considering these results, we can conclude that DeepFault is effective in 
synthesizing highly adversarial inputs 
(cf. Table~\ref{table:models_validation_avg})
that closely resemble their original counterparts. 

\begin{table*}[t]
	\caption{Distance between synthesized and original inputs. The values shown 
		represent 
		minimal perturbation to the original inputs ($\!<\!5\%$ for MNIST and 
		$\!<\!1\%$ for 
		CIFAR-10).}
	\label{table:similarity}
	\vspace*{-3mm}
	\begin{minipage}{.5\textwidth}
		\centering
		\scalebox{0.88}{
			\begin{tabular}{|l|l|l|l|l|l|l|}
				\hline
				\multicolumn{1}{|c|}{\textbf{$k$}}& 
				\multicolumn{3}{c|}{\textbf{MNIST}} & 
				\multicolumn{3}{c|}{\textbf{CIFAR}} \\ \cline{2-7}
				{\scriptsize MNIST(CIFAR)}& \textbf{$L_1$} & \textbf{$L_2$} & 
				\textbf{$L_\infty$} & 
				\textbf{$L_1$} & \textbf{$L_2$} & \textbf{$L_\infty$} \\ \hline
				\textbf{1(10)} & 41.4 & 2.0 & 0.1 & 179.07 & 7216.6 & 15.46 
				\\
				\textbf{2(20)} & 41.2 & 1.99 & 0.1 & 144.95 & 5897.4 & 12.45 
				\\
				\textbf{3(30)} & 40.9 & 1.98 & 0.1 & 124.61 & 5073.9 & 10.67 
				\\
				\textbf{5(40)} & 40.7 & 1.97 & 0.1 & 113.45 & 4579.2 & 9.89 
				\\
				\textbf{10(50)} & 40.3 & 1.96 & 0.1 & 104.72 & 4273  & 9.24 
				\\ \hline
		\end{tabular}}
	\end{minipage}%
	\begin{minipage}{.5\textwidth}
		\centering
		\vspace*{-3mm}
		\scalebox{0.88}{
			\begin{tabular}{|l|c|c|c|c|c|c|}
				\hline
				\textbf{Susp.}
				& \multicolumn{3}{c|}{\textbf{MNIST}} & 
				\multicolumn{3}{c|}{\textbf{CIFAR}} \\ \cline{2-7}
				\textbf{measure}& \textbf{$L_1$} & \textbf{$L_2$} & 
				\textbf{$L_\infty$} 
				& 
				\textbf{$L_1$} & \textbf{$L_2$} & \textbf{$L_\infty$} \\ \hline
				\textbf{Tarantula} & 40.3 & 1.97 & 0.1 & 180.23 & 6575.6 & 
				19.41 \\ 
				\textbf{Ochiai} & 41.0 & 1.98 & 0.1 & 110.45 & 4825.3 & 7.84 \\
				\textbf{DStar} & 41.5 & 1.99 & 0.1 & 109.4 & 4823.2 & 7.39 \\
				\textbf{Random} & 39.2 & 1.92 & 0.1 & 121.73 & 4988.1 & 11.63 
				\\ \hline
		\end{tabular}}
	\end{minipage}%
\end{table*}

\vspace*{-4mm}
\subsubsection{\textbf{RQ5 (Increasing Activations).}}

\begin{wraptable}{R}{0.6\textwidth}
	\centering
	\vspace*{-3mm}
	\caption{Effectiveness of \textit{suspiciousness-guided input synthesis} 
	algorithm to increase activations values of suspicious neurons.}
	\label{table:effect_avg}
	\renewcommand{\arraystretch}{0.9}
	\begin{tabular}{|l|c|c|c|c|c|c|}
		\hline
		&\multicolumn{5}{c|}{$k$: MNIST(CIFAR)}\\
		\textbf{Datasets} 
		&\textbf{1(10)}
		&\textbf{2(20)} 
		&\textbf{3(30)} 
		&\textbf{5(40)}
		&\textbf{10(50)}\\ \hline
		\textbf{MNIST} 		
		& 98\% & 99\% & 97\% & 97\% & 91\% \\ 
		\textbf{CIFAR} 
		& 91\% & 92\% & 90\% & 89\% & 88\% \\ \hline
	\end{tabular}
	\vspace*{-8mm}
\end{wraptable}

We studied the activation values of suspicious neurons identified by DeepFault 
to examine whether the synthesized inputs increase the values of these 
neurons.  
The gradients of suspicious neurons used in our suspiciousness-guided input 
synthesis algorithm might be conflicting and a global increase in all 
suspicious neurons' values might not be feasible. 
This can occur 
if some neurons' 
gradients are negative, indicating a decrease in an input feature's value, 
whereas other gradients are positive and require to increase the value of 
the same feature. 
Table~\ref{table:effect_avg} shows the percentage of suspicious neurons $k$, 
averaged over all suspiciousness measures for all considered MNIST and CIFAR-10 
models from Table~\ref{table:dnns}, 
whose values were increased by the inputs synthesized by DeepFault. 
For MNIST models, DeepFault synthesized inputs that increase the suspicious 
neurons' values with success at least 97$\%$ for 
$k\!\in\!\{\!1,\!2,\!3,\!5\!\}$, while the average effectiveness for CIFAR 
models is 90\%.
These results show the effectiveness of our suspiciousness-guided input 
synthesis algorithm in generating inputs that increase the activation values 
of suspicious neurons (see Appendix).

\vspace*{-6mm}
\subsubsection{\textbf{RQ6 (Performance).}}
We measured the performance of Algorithm~\ref{alg:syn} to synthesize new 
inputs (Table~\ref{table:performance} in Appendix). 
The average time required to synthesize a single input for MNIST and CIFAR 
models is $1s$ and $24.3s$, respectively.
As expected, the performance of the algorithm depends on the number 
of suspicious neurons ($k$), the distribution of those neurons over the DNN and 
its architecture. 
For CIFAR models, for instance, the execution time per input ranges between 
$3s$ ($k\!=\!10$) and $48s$ ($k\!=\!50$). 
We also confirmed empirically that more time is taken to synthesize an input if 
the suspicious neurons are in deeper hidden layers.

\subsection{Threats to Validity \label{sec:threats}}
\vspace*{-2mm}

\textbf{Construct validity} threats might be due to the adopted experimental 
methodology 
including the selected datasets and DNN models. 
To mitigate this threat, we used widely studied public datasets 
(MNIST~\cite{lecun1998mnist} and CIFAR-10~\cite{cifar_model}), and applied 
DeepFault to multiple DNN models of different architectures 
with competitive prediction accuracies (cf. Table~\ref{table:dnns}).
Also, we mitigate threats related to the identification of suspicious 
neurons (Algorithm~\ref{alg:identification}) by adapting suspiciousness 
measures from the fault localisation domain in software 
engineering~\cite{wong2016survey}. 

\noindent
\textbf{Internal validity} threats might occur
when establishing the ability of DeepFault to synthesize new inputs that exercise 
the identified suspicious neurons.
 To mitigate this threat, we used various distance metrics 
 to confirm that the synthesized inputs are 
 close to the original inputs and 
 similar to the inputs synthesized by a random 
 strategy.
Another threat could be that the suspiciousness measures employed by DeepFault 
accidentally outperform the random strategy. 
To mitigate this threat, we reported the results of the random strategy over 
five independent runs per experiment. 
Also, we ensured that the distribution of the randomly selected suspicious 
neurons resembles the distribution of neurons identified by 
DeepFault suspiciousness measures. 
We also used the non-parametric statistical test Mann-Whitney to 
check for significant difference in the performance of DeepFault instances and 
random with a 95\% confidence level.
\
\\
\noindent
\textbf{External validity} threats might exist if DeepFault cannot access the 
internal DNN structure to assemble the hit spectrums of neurons and 
establish their suspiciousness.
We limit this threat by developing DeepFault using the open-source frameworks Keras and Tensorflow which enable whitebox DNN analysis.
We also examined various spectrum-based suspiciousness measures, but other measures can be investigated~\cite{wong2016survey}. 
We further reduce the risk that DeepFault might be difficult to use in 
practice by validating it against several DNN instances trained on two widely-used datasets.
However, more experiments are needed to assess the applicability of DeepFault
in domains and networks with characteristics 
different from those used in our evaluation (e.g., LSTM and Capsules networks~\cite{sabour2017dynamic}). 

\vspace*{-2mm}
\section{Related Work\label{sec:relatedWork}}
\vspace*{-3mm}

\textbf{DNN Testing and Verification.}
The inability of blackbox DNN testing
to provide insights about the internal 
neuron activity and enable identification of corner-case inputs that expose 
unexpected network behaviour~\cite{goodfellow2017challenge}, 
urged researchers 
to leverage whitebox testing techniques from software 
engineering~\cite{pei2017deepxplore,OG18,SHK18,KFY18,MJX18}. 
DeepXplore~\cite{pei2017deepxplore} uses a differential algorithm to 
generate inputs that increase neuron coverage. 
DeepGauge~\cite{MJX18} introduces multi-granularity coverage criteria for 
effective test synthesis. 
Other research proposes testing criteria and techniques inspired by metamorphic 
testing~\cite{TPSR18}, combinatorial testing~\cite{MZXLLZW18}, 
mutation testing$\!$~\cite{MZSXJ18}, 
MC/DC$\!$~\cite{SHK18},
symbolic execution$\!$~\cite{GWZPK18} and concolic testing$\!$~\cite{SWRHKK18}. 

Formal DNN verification 
aims at providing guarantees for trustworthy DNN operation~\cite{Huang18}.
Abstraction refinement is used in~\cite{pulina2010abstraction} to verify safety 
properties of small neural networks with sigmoid activation functions, while 
AI$^2$~\cite{GMDTCV18} employs abstract interpretation
to verify similar properties.
Reluplex~\cite{katz2017reluplex} is an SMT-based approach that verifies safety 
and robustness of DNNs with ReLUs, 
and DeepSafe~\cite{gopinath2017deepsafe} uses Reluplex to identify safe 
regions in the input space. 
DLV~\cite{wicker2018feature} can verify local DNN robustness 
given a set of user-defined manipulations.

DeepFault adopts spectrum-based fault localisation techniques to systematically 
identify suspicious neurons and uses these neurons to synthesize new inputs, 
which is mostly orthogonal to existing research on DNN testing and verification.

\vspace*{1mm}
\noindent
\textbf{Adversarial Deep Learning.}
Recent studies have shown that DNNs are vulnerable to adversarial 
examples~\cite{szegedy2013intriguing} and proposed search algorithms 
~\cite{carlini2017towards,papernot2016limitations,moosavi2016deepfool,nguyen2015deep},
based on gradient descent or optimisation techniques, for generating 
adversarial inputs that have a minimal difference to their 
original versions and force the DNN to exhibit erroneous behaviour. 
These types of adversarial examples have been shown to exist in the physical 
world too~\cite{kurakin2016adversarial}. 
The identification of and protection against these adversarial attacks, 
is another active area of research~\cite{tramer2017ensemble,papernot2016distillation}. 
DeepFault is similar to these approaches since 
it uses the identified suspicious neurons to synthesize perturbed inputs which 
as we have demonstrated in Section~\ref{sec:evaluation} are adversarial. 
Extending DeepFault to support the synthesis of adversarial inputs using these 
adversarial search algorithms is part of our future work.

\vspace*{1mm}
\noindent
\textbf{Fault Localization in Traditional Software.}
Fault localization is widely studied in many software 
engineering areas including including software debugging~\cite{Parnin2011ISSTA}, program 
repair~\cite{Goues2012TSE} and failure 
reproduction~\cite{Jin2012BugRedux,Jin2013}. 
The research focus in fault localization is the development of 
identification methods and suspiciousness measures that isolate the root 
causes of software failures with reduced engineering 
effort~\cite{pearson2017evaluating}.  
The most notable fault localization methods are 
spectrum-based~\cite{landsberg2015evaluation,landsberg2018optimising,wong2014dstar,jones2005empirical,abreu2009practical},
slice-based~\cite{Wong2006Slicing} and model-based~\cite{Mayer2008ModelFault}. 
Threats to the value of empirical evaluations of spectrum-based fault 
localisation 
are studied in~\cite{Steimann2013ISSTA}, while the 
theoretical analyses in~\cite{Yoo2017TOSEM,Xie2013TOSEM} set a 
formal foundation about desirable formal properties that suspiciousness 
measures should have.
We refer interested readers to a recent comprehensive survey on fault 
localization~\cite{wong2016survey}.

\vspace*{-2mm}
\section{Conclusion \label{sec:conclusion}}
\vspace*{-4mm}
The potential deployment of DNNs in safety-critical applications  introduces unacceptable risks.
To reduce these risks to acceptable levels, DNNs should be tested thoroughly. 
We contribute in this effort, by introducing DeepFault, the first 
fault localization-based whitebox testing approach for DNNs. 
DeepFault \emph{analyzes} pre-trained DNNs, given a specific 
test set, to establish the hit spectrum of each neuron, \emph{identifies 
suspicious neurons} by employing suspiciousness measures and \emph{synthesizes} 
new inputs that increase the activation values of the suspicious neurons. 
Our empirical evaluation on the widely-used MNIST and CIFAR-10 datasets shows 
that DeepFault can identify neurons which can be held responsible for 
inadequate performance. 
DeepFault can also synthesize new inputs, which closely resemble the original 
inputs, are highly adversarial and exercise the identified suspicious neurons. 
In future work, we plan to evaluate DeepFault on other DNNs and datasets, to 
improve the suspiciousness-guided synthesis algorithm and to extend the 
synthesis of adversarial inputs~\cite{papernot2016limitations}. 
We will also explore techniques to repair the identified suspicious neurons, thus enabling to reason about the safety of DNNs and support safety case generation~\cite{calinescu2018engineering,kelly1999arguing}.

\newpage
\section*{Appendix}

	
	\subsection*{Fault Localization in Software Engineering}
	\vspace*{-5mm}
	\vspace*{-10mm}
	
	\subsection*{Additional Material for RQ5\label{app:valueincrease}}
	\vspace*{-10mm}
	\begin{table}[h!]
		\caption{Ratios of increases in activations of $k^\prime$ suspicious 
		neurons with inputs that are synthesized to increase activations of $k$ 
		neurons where $k^\prime \geq k$ in MNIST networks. 
		Synthesized inputs, most of the time, increase the activation values 
		of the top $k^\prime$ neurons in addition to increasing the values of 
		the top $k$ suspicious neurons.} 
		\label{table:auto_inc_susp_mnist}
		\centering
		\begin{tabular}{|c|c|c|c|c|c|c|c|c|c|c|c|c|c|c|c|}
			\hline
			& \multicolumn{5}{c|}{\textbf{Tarantula}} & 
			\multicolumn{5}{c|}{\textbf{Ochiai}} & 
			\multicolumn{5}{c|}{\textbf{D$^*$}} \\ \hline
			\backslashbox{$k$}{$k^\prime$} & \textbf{1} & \textbf{2} & 
			\textbf{3} & 
			\textbf{5} & \textbf{10} & \textbf{1} & \textbf{2} & \textbf{3} & 
			\textbf{5} & \textbf{10} & \textbf{1} & \textbf{2} & \textbf{3} & 
			\textbf{5} & \textbf{10} \\ \hline
			\textbf{1} & 1.0 & 0.98 & 0.94 & 0.90 & 0.78 & 1.0 & 1.0 & 0.96 & 
			0.97 
			& 0.90 & 1.0 & 0.95 & 0.96 & 0.97 & 0.84 \\ \hline
			\textbf{2} & \cellcolor[HTML]{333333} & 1.0 & 0.98 & 0.91 & 0.81 & 
			\cellcolor[HTML]{333333}{\color[HTML]{333333} } & 1.0 & 0.99 & 0.99 
			& 
			0.89 & \cellcolor[HTML]{333333} & 1.0 & 0.98 & 0.98 & 0.87 \\ \hline
			\textbf{3} & \cellcolor[HTML]{333333} & \cellcolor[HTML]{333333} & 
			0.98 
			& 0.97 & 0.87 & \cellcolor[HTML]{333333}{\color[HTML]{333333} } & 
			\cellcolor[HTML]{333333} & 1.0 & 0.99 & 0.90 & 
			\cellcolor[HTML]{333333} 
			& \cellcolor[HTML]{333333} & 1.0 & 1.0 & 0.91 \\ \hline
			\textbf{5} & \cellcolor[HTML]{333333} & \cellcolor[HTML]{333333} & 
			\cellcolor[HTML]{333333} & 0.98 & 0.89 & 
			\cellcolor[HTML]{333333}{\color[HTML]{333333} } & 
			\cellcolor[HTML]{333333} & \cellcolor[HTML]{333333} & 0.98 & 0.91 & 
			\cellcolor[HTML]{333333} & \cellcolor[HTML]{333333} & 
			\cellcolor[HTML]{333333} & 1.0 & 0.90 \\ \hline
			\textbf{10} & \cellcolor[HTML]{333333} & \cellcolor[HTML]{333333} & 
			\cellcolor[HTML]{333333} & \cellcolor[HTML]{333333} & 0.91 & 
			\cellcolor[HTML]{333333}{\color[HTML]{333333} } & 
			\cellcolor[HTML]{333333} & \cellcolor[HTML]{333333} & 
			\cellcolor[HTML]{333333} & 0.97 & \cellcolor[HTML]{333333} & 
			\cellcolor[HTML]{333333} & \cellcolor[HTML]{333333} & 
			\cellcolor[HTML]{333333} & 0.94 \\ \hline
		\end{tabular}
	\end{table}
	
	\vspace*{-10mm}
	\begin{table}[h!]
		\caption{Ratios of increases in activations of $k^\prime$ suspicious 
		neurons with inputs that are synthesized to increase activations of $k$ 
		neurons where $k^\prime \geq k$ in CIFAR networks. 
		Synthesized inputs, most of the time (except from D$^*$ case), 
		increase the activation values of the top $k^\prime$ neurons in 
		addition to increasing the values of the top $k$ suspicious neurons.}
		\label{table:auto_inc_susp_cifar}
		\centering
		\begin{tabular}{|c|c|c|c|c|c|c|c|c|c|c|c|c|c|c|c|}
			\hline
			& \multicolumn{5}{c|}{\textbf{Tarantula}} & 
			\multicolumn{5}{c|}{\textbf{Ochiai}} & 
			\multicolumn{5}{c|}{\textbf{D$^*$}} \\ \hline
			\backslashbox{$k$}{$k^\prime$} & \textbf{10} & \textbf{20} & 
			\textbf{30} & \textbf{40} & \textbf{50} & \textbf{10} & \textbf{20} 
			& 
			\textbf{30} & \textbf{40} & \textbf{50} & \textbf{10} & \textbf{20} 
			& 
			\textbf{30} & \textbf{40} & \textbf{50} \\ \hline
			\textbf{10} & 0.95 & 0.91 & 0.92 & 0.90 & 0.89 & 0.95 & 0.92 & 0.92 
			& 
			0.91 & 0.89 & 0.81 & 0.61 & 0.56 & 0.32 & 0.15 \\ \hline
			\textbf{20} & \cellcolor[HTML]{333333} & 0.96 & 0.94 & 0.92 & 0.90 
			& 
			\cellcolor[HTML]{333333}{\color[HTML]{333333} } & 0.97 & 0.95 & 
			0.94 & 
			0.91 & \cellcolor[HTML]{333333} & 0.78 & 0.49 & 0.31 & 0.15 \\ 
			\hline
			\textbf{30} & \cellcolor[HTML]{333333} & \cellcolor[HTML]{333333} & 
			0.94 & 0.92 & 0.91  & \cellcolor[HTML]{333333}{\color[HTML]{333333} 
			} & 
			\cellcolor[HTML]{333333} & 0.92 & 0.91 & 0.90 & 
			\cellcolor[HTML]{333333} & \cellcolor[HTML]{333333} & 0.79 & 0.33 & 
			0.15 \\ \hline
			\textbf{40} & \cellcolor[HTML]{333333} & \cellcolor[HTML]{333333} & 
			\cellcolor[HTML]{333333} & 0.95 & 0.93 & 
			\cellcolor[HTML]{333333}{\color[HTML]{333333} } & 
			\cellcolor[HTML]{333333} & \cellcolor[HTML]{333333} & 0.97 & 0.93 & 
			\cellcolor[HTML]{333333} & \cellcolor[HTML]{333333} & 
			\cellcolor[HTML]{333333} & 0.65 & 0.17 \\ \hline
			\textbf{50} & \cellcolor[HTML]{333333} & \cellcolor[HTML]{333333} & 
			\cellcolor[HTML]{333333} & \cellcolor[HTML]{333333} & 0.94 & 
			\cellcolor[HTML]{333333}{\color[HTML]{333333} } & 
			\cellcolor[HTML]{333333} & \cellcolor[HTML]{333333} & 
			\cellcolor[HTML]{333333} & 0.91 & \cellcolor[HTML]{333333} & 
			\cellcolor[HTML]{333333} & \cellcolor[HTML]{333333} & 
			\cellcolor[HTML]{333333} & 0.7 \\ \hline
		\end{tabular}
	\end{table}

	\newpage
	\subsection*{Additional Material for RQ2}\label{app:comparison}
	\vspace*{-9mm}
	\begin{table}[!ht]
		\caption{Statistical significance evaluation of techniques when 
		compared to 
			each other on MNIST networks. A checkmark is put when a technique 
			on the row 
			has a significantly better loss/accuracy value than the technique 
			on the 
			column for given network and $k$ parameter ($k$ stands for number 
			of 
			suspicious neurons).}
		\label{table:stat_sign_rq2_mnist}
		\renewcommand{\arraystretch}{0.83}
		\vspace*{-3mm}
		\footnotesize
		\centering
		\begin{tabular}{|c|c|c|c|c|c|c|c|c|}
			\hline
			& && \multicolumn{2}{c|}{\textbf{Tarantula}} & 
			\multicolumn{2}{c|}{\textbf{Ochiai}} & 
			\multicolumn{2}{c|}{\textbf{D$^*$}} \\ \hline
			\textbf{Technique} & \textbf{Network} & $k$ & \textbf{Loss} & 
			\textbf{Accuracy} & \textbf{Loss} & \textbf{Accuracy} & 
			\textbf{Loss} & 
			\textbf{Accuracy} \\ \hline
			&  & \textbf{1} & \cellcolor[HTML]{333333} & 
			\cellcolor[HTML]{333333} 
			&  &  &  &  \\  
			&  & \textbf{2} & \cellcolor[HTML]{333333} & 
			\cellcolor[HTML]{333333} 
			&  & \checkmark &  &  \\  
			&  & \textbf{3} & \cellcolor[HTML]{333333} & 
			\cellcolor[HTML]{333333} 
			&  &  &  &  \\  
			&  & \textbf{5} & \cellcolor[HTML]{333333} & 
			\cellcolor[HTML]{333333} 
			&  &  &  &  \\  
			& \multirow{-5}{*}{\textbf{MNIST\_1}} & \textbf{10} & 
			\cellcolor[HTML]{333333} & \cellcolor[HTML]{333333} &  &  &  &  \\ 
			\cline{2-9} 
			&  & \textbf{1} & \cellcolor[HTML]{333333} & 
			\cellcolor[HTML]{333333} 
			&  &  &  &  \\  
			&  & \textbf{2} & \cellcolor[HTML]{333333} & 
			\cellcolor[HTML]{333333} 
			&  &  &  &  \\  
			&  & \textbf{3} & \cellcolor[HTML]{333333} & 
			\cellcolor[HTML]{333333} 
			&  &  &  &  \\  
			&  & \textbf{5} & \cellcolor[HTML]{333333} & 
			\cellcolor[HTML]{333333} 
			&  &  &  &  \\  
			& \multirow{-5}{*}{\textbf{MNIST\_2}} & \textbf{10} & 
			\cellcolor[HTML]{333333} & \cellcolor[HTML]{333333} &  &  &  &  \\ 
			\cline{2-9} 
			&  & \textbf{1} & \cellcolor[HTML]{333333} & 
			\cellcolor[HTML]{333333} 
			&  &  &  &  \\  
			&  & \textbf{2} & \cellcolor[HTML]{333333} & 
			\cellcolor[HTML]{333333} 
			&  &  &  &  \\  
			&  & \textbf{3} & \cellcolor[HTML]{333333} & 
			\cellcolor[HTML]{333333} 
			&  &  &  &  \\  
			&  & \textbf{5} & \cellcolor[HTML]{333333} & 
			\cellcolor[HTML]{333333} 
			&  &  &  &  \\  
			\multirow{-15}{*}{\textbf{Tarantula}} & 
			\multirow{-5}{*}{\textbf{MNIST\_3}} & \textbf{10} & 
			\cellcolor[HTML]{333333} & \cellcolor[HTML]{333333} &  &  &  &  \\ 
			\hline
			&  & \textbf{1} & \checkmark & \checkmark & 
			\cellcolor[HTML]{333333} & 
			\cellcolor[HTML]{333333} & \checkmark &  \\  
			&  & \textbf{2} & \checkmark &  & \cellcolor[HTML]{333333} & 
			\cellcolor[HTML]{333333} & \checkmark &  \\  
			&  & \textbf{3} & \checkmark &  & \cellcolor[HTML]{333333} & 
			\cellcolor[HTML]{333333} &  &  \\  
			&  & \textbf{5} & \checkmark &  & \cellcolor[HTML]{333333} & 
			\cellcolor[HTML]{333333} &  &  \\  
			& \multirow{-5}{*}{\textbf{MNIST\_1}} & \textbf{10} & \checkmark &  
			& 
			\cellcolor[HTML]{333333} & \cellcolor[HTML]{333333} &  &  \\ 
			\cline{2-9} 
			&  & \textbf{1} &  &  & \cellcolor[HTML]{333333} & 
			\cellcolor[HTML]{333333} &  & \checkmark \\  
			&  & \textbf{2} &  &  & \cellcolor[HTML]{333333} & 
			\cellcolor[HTML]{333333} &  &  \\  
			&  & \textbf{3} & \checkmark & \checkmark & 
			\cellcolor[HTML]{333333} & 
			\cellcolor[HTML]{333333} &  &  \\  
			&  & \textbf{5} & \checkmark &  & \cellcolor[HTML]{333333} & 
			\cellcolor[HTML]{333333} &  &  \\  
			& \multirow{-5}{*}{\textbf{MNIST\_2}} & \textbf{10} &  & \checkmark 
			& 
			\cellcolor[HTML]{333333} & \cellcolor[HTML]{333333} &  &  \\ 
			\cline{2-9} 
			&  & \textbf{1} &  &  & \cellcolor[HTML]{333333} & 
			\cellcolor[HTML]{333333} & \checkmark &  \\  
			&  & \textbf{2} &  &  & \cellcolor[HTML]{333333} & 
			\cellcolor[HTML]{333333} &  &  \\  
			&  & \textbf{3} &  &  & \cellcolor[HTML]{333333} & 
			\cellcolor[HTML]{333333} &  &  \\  
			&  & \textbf{5} &  &  & \cellcolor[HTML]{333333} & 
			\cellcolor[HTML]{333333} &  & \checkmark \\  
			\multirow{-15}{*}{\textbf{Ochiai}} & 
			\multirow{-5}{*}{\textbf{MNIST\_3}} & \textbf{10} &  &  & 
			\cellcolor[HTML]{333333} & \cellcolor[HTML]{333333} & \checkmark &  
			\\ 
			\hline
			&  & \textbf{1} &  &  &  &  & \cellcolor[HTML]{333333} & 
			\cellcolor[HTML]{333333} \\  
			&  & \textbf{2} &  &  &  &  & \cellcolor[HTML]{333333} & 
			\cellcolor[HTML]{333333} \\  
			&  & \textbf{3} & \checkmark &  &  &  & \cellcolor[HTML]{333333} & 
			\cellcolor[HTML]{333333} \\  
			&  & \textbf{5} & \checkmark &  &  &  & \cellcolor[HTML]{333333} & 
			\cellcolor[HTML]{333333} \\  
			& \multirow{-5}{*}{\textbf{MNIST\_1}} & \textbf{10} & \checkmark &  
			&  
			&  & \cellcolor[HTML]{333333} & \cellcolor[HTML]{333333} \\ 
			\cline{2-9} 
			&  & \textbf{1} &  &  &  &  & \cellcolor[HTML]{333333} & 
			\cellcolor[HTML]{333333} \\  
			&  & \textbf{2} &  &  &  &  & \cellcolor[HTML]{333333} & 
			\cellcolor[HTML]{333333} \\  
			&  & \textbf{3} &  & \checkmark &  &  & \cellcolor[HTML]{333333} & 
			\cellcolor[HTML]{333333} \\  
			&  & \textbf{5} &  &  &  &  & \cellcolor[HTML]{333333} & 
			\cellcolor[HTML]{333333} \\  
			& \multirow{-5}{*}{\textbf{MNIST\_2}} & \textbf{10} & \checkmark & 
			\checkmark &  &  & \cellcolor[HTML]{333333} & 
			\cellcolor[HTML]{333333} 
			\\ \cline{2-9} 
			&  & \textbf{1} &  &  &  &  & \cellcolor[HTML]{333333} & 
			\cellcolor[HTML]{333333} \\  
			&  & \textbf{2} & \checkmark &  &  &  & \cellcolor[HTML]{333333} & 
			\cellcolor[HTML]{333333} \\  
			&  & \textbf{3} & \checkmark &  &  &  & \cellcolor[HTML]{333333} & 
			\cellcolor[HTML]{333333} \\  
			&  & \textbf{5} & \checkmark &  &  &  & \cellcolor[HTML]{333333} & 
			\cellcolor[HTML]{333333} \\  
			\multirow{-15}{*}{\textbf{D$^*$}} & 
			\multirow{-5}{*}{\textbf{MNIST\_3}} 
			& \textbf{10} & \checkmark &  &  &  & \cellcolor[HTML]{333333} & 
			\cellcolor[HTML]{333333} \\ \hline
		\end{tabular}
	\end{table}

	\begin{table}[!ht]
		\caption{Statistical significance evaluation of techniques when 
		compared to 
			each other on CIFAR networks. A checkmark is put when a technique 
			on the 
			row has a significantly better loss/accuracy value than the 
			technique on 
			the column for given network and $k$ parameter 
			($k$ stands for number of suspicious neurons).}
		\label{table:stat_sign_rq2_cifar}
		\renewcommand{\arraystretch}{0.83}
		\vspace*{-3mm}
		\footnotesize
		\centering
		\begin{tabular}{|c|c|c|c|c|c|c|c|c|}
			\hline
			& &  & \multicolumn{2}{c|}{\textbf{Tarantula}} & 
			\multicolumn{2}{c|}{\textbf{Ochiai}} & 
			\multicolumn{2}{c|}{\textbf{D$^*$}} \\ \hline
			\textbf{Technique} & \textbf{Network} & $k$ & \textbf{Loss} & 
			\textbf{Accuracy} & \textbf{Loss} & \textbf{Accuracy} & 
			\textbf{Loss} & 
			\textbf{Accuracy} \\ \hline
			&  & \textbf{10} & \cellcolor[HTML]{333333} & 
			\cellcolor[HTML]{333333} 
			&  &  & \checkmark & \checkmark \\  
			&  & \textbf{20} & \cellcolor[HTML]{333333} & 
			\cellcolor[HTML]{333333} 
			&  &  & \checkmark & \checkmark \\  
			&  & \textbf{30} & \cellcolor[HTML]{333333} & 
			\cellcolor[HTML]{333333} 
			&  &  & \checkmark & \checkmark \\  
			&  & \textbf{40} & \cellcolor[HTML]{333333} & 
			\cellcolor[HTML]{333333} 
			&  & \checkmark & \checkmark & \checkmark \\  
			& \multirow{-5}{*}{\textbf{CIFAR\_1}} & \textbf{50} & 
			\cellcolor[HTML]{333333} & \cellcolor[HTML]{333333} &  &  & 
			\checkmark 
			& \checkmark \\ \cline{2-9} 
			&  & \textbf{10} & \cellcolor[HTML]{333333} & 
			\cellcolor[HTML]{333333} 
			&  &  & \checkmark & \checkmark \\  
			&  & \textbf{20} & \cellcolor[HTML]{333333} & 
			\cellcolor[HTML]{333333} 
			& \checkmark &  & \checkmark & \checkmark \\  
			&  & \textbf{30} & \cellcolor[HTML]{333333} & 
			\cellcolor[HTML]{333333} 
			& \checkmark &  & \checkmark & \checkmark \\  
			&  & \textbf{40} & \cellcolor[HTML]{333333} & 
			\cellcolor[HTML]{333333} 
			& \checkmark & \checkmark & \checkmark & \checkmark \\  
			& \multirow{-5}{*}{\textbf{CIFAR\_2}} & \textbf{50} & 
			\cellcolor[HTML]{333333} & \cellcolor[HTML]{333333} & \checkmark & 
			\checkmark & \checkmark & \checkmark \\ \cline{2-9} 
			&  & \textbf{10} & \cellcolor[HTML]{333333} & 
			\cellcolor[HTML]{333333} 
			& \checkmark & \checkmark & \checkmark & \checkmark \\  
			&  & \textbf{20} & \cellcolor[HTML]{333333} & 
			\cellcolor[HTML]{333333} 
			& \checkmark & \checkmark & \checkmark & \checkmark \\  
			&  & \textbf{30} & \cellcolor[HTML]{333333} & 
			\cellcolor[HTML]{333333} 
			& \checkmark & \checkmark & \checkmark & \checkmark \\  
			&  & \textbf{40} & \cellcolor[HTML]{333333} & 
			\cellcolor[HTML]{333333} 
			& \checkmark & \checkmark & \checkmark & \checkmark \\  
			\multirow{-15}{*}{\textbf{Tarantula}} & 
			\multirow{-5}{*}{\textbf{CIFAR\_3}} & \textbf{50} & 
			\cellcolor[HTML]{333333} & \cellcolor[HTML]{333333} & \checkmark & 
			\checkmark & \checkmark & \checkmark \\ \hline
			&  & \textbf{10} &  &  & \cellcolor[HTML]{333333} & 
			\cellcolor[HTML]{333333} & \checkmark & \checkmark \\  
			&  & \textbf{20} &  &  & \cellcolor[HTML]{333333} & 
			\cellcolor[HTML]{333333} & \checkmark & \checkmark \\  
			&  & \textbf{30} &  &  & \cellcolor[HTML]{333333} & 
			\cellcolor[HTML]{333333} & \checkmark & \checkmark \\  
			&  & \textbf{40} &  &  & \cellcolor[HTML]{333333} & 
			\cellcolor[HTML]{333333} & \checkmark & \checkmark \\  
			& \multirow{-5}{*}{\textbf{CIFAR\_1}} & \textbf{50} &  &  & 
			\cellcolor[HTML]{333333} & \cellcolor[HTML]{333333} & \checkmark & 
			\checkmark \\ \cline{2-9} 
			&  & \textbf{10} &  &  & \cellcolor[HTML]{333333} & 
			\cellcolor[HTML]{333333} & \checkmark & \checkmark \\  
			&  & \textbf{20} &  &  & \cellcolor[HTML]{333333} & 
			\cellcolor[HTML]{333333} & \checkmark & \checkmark \\  
			&  & \textbf{30} &  &  & \cellcolor[HTML]{333333} & 
			\cellcolor[HTML]{333333} & \checkmark & \checkmark \\  
			&  & \textbf{40} &  &  & \cellcolor[HTML]{333333} & 
			\cellcolor[HTML]{333333} & \checkmark & \checkmark \\  
			& \multirow{-5}{*}{\textbf{CIFAR\_2}} & \textbf{50} &  &  & 
			\cellcolor[HTML]{333333} & \cellcolor[HTML]{333333} & \checkmark & 
			\checkmark \\ \cline{2-9} 
			&  & \textbf{10} &  &  & \cellcolor[HTML]{333333} & 
			\cellcolor[HTML]{333333} &  &  \\  
			&  & \textbf{20} &  &  & \cellcolor[HTML]{333333} & 
			\cellcolor[HTML]{333333} &  &  \\  
			&  & \textbf{30} &  &  & \cellcolor[HTML]{333333} & 
			\cellcolor[HTML]{333333} &  &  \\  
			&  & \textbf{40} &  &  & \cellcolor[HTML]{333333} & 
			\cellcolor[HTML]{333333} &  &  \\  
			\multirow{-15}{*}{\textbf{Ochiai}} & 
			\multirow{-5}{*}{\textbf{CIFAR\_3}} & \textbf{50} &  &  & 
			\cellcolor[HTML]{333333} & \cellcolor[HTML]{333333} &  &  \\ \hline
			&  & \textbf{10} &  &  &  &  & \cellcolor[HTML]{333333} & 
			\cellcolor[HTML]{333333} \\  
			&  & \textbf{20} &  &  &  &  & \cellcolor[HTML]{333333} & 
			\cellcolor[HTML]{333333} \\  
			&  & \textbf{30} &  &  &  &  & \cellcolor[HTML]{333333} & 
			\cellcolor[HTML]{333333} \\  
			&  & \textbf{40} &  &  &  &  & \cellcolor[HTML]{333333} & 
			\cellcolor[HTML]{333333} \\  
			& \multirow{-5}{*}{\textbf{CIFAR\_1}} & \textbf{50} &  &  &  &  & 
			\cellcolor[HTML]{333333} & \cellcolor[HTML]{333333} \\ \cline{2-9} 
			&  & \textbf{10} &  &  &  &  & \cellcolor[HTML]{333333} & 
			\cellcolor[HTML]{333333} \\  
			&  & \textbf{20} &  &  &  &  & \cellcolor[HTML]{333333} & 
			\cellcolor[HTML]{333333} \\  
			&  & \textbf{30} &  &  &  &  & \cellcolor[HTML]{333333} & 
			\cellcolor[HTML]{333333} \\  
			&  & \textbf{40} &  &  &  &  & \cellcolor[HTML]{333333} & 
			\cellcolor[HTML]{333333} \\  
			& \multirow{-5}{*}{\textbf{CIFAR\_2}} & \textbf{50} &  &  &  &  & 
			\cellcolor[HTML]{333333} & \cellcolor[HTML]{333333} \\ \cline{2-9} 
			&  & \textbf{10} &  &  &  &  & \cellcolor[HTML]{333333} & 
			\cellcolor[HTML]{333333} \\  
			&  & \textbf{20} &  &  &  &  & \cellcolor[HTML]{333333} & 
			\cellcolor[HTML]{333333} \\  
			&  & \textbf{30} &  &  &  &  & \cellcolor[HTML]{333333} & 
			\cellcolor[HTML]{333333} \\  
			&  & \textbf{40} &  &  &  &  & \cellcolor[HTML]{333333} & 
			\cellcolor[HTML]{333333} \\  
			\multirow{-15}{*}{\textbf{D$^*$}} & 
			\multirow{-5}{*}{\textbf{CIFAR\_3}} 
			& \textbf{50} &  &  &  &  & \cellcolor[HTML]{333333} & 
			\cellcolor[HTML]{333333} \\ \hline
		\end{tabular}
	\end{table}

	\newpage \ \newpage
	\subsection*{Additional Material for RQ6}
	\vspace*{-10mm}
	\begin{table}[h!]
		\centering
		\caption{Performance measurement of input perturbation in DeepFault. 
			(mm:ss)}
		\vspace*{-3mm}
		\label{table:performance}
		\begin{tabular}{|l|c|c|c|c|c|c|}
			\hline
			\textbf{DNNs} & \textbf{Susp. Measure} & \textbf{$k$=1(10)} & 
			\textbf{$k$=2(20)} & \textbf{$k$=3(30)} & \textbf{$k$=5(40)} & 
			\textbf{$k$=10(50)} \\ \hline
			\multirow{3}{*}{\textbf{MNIST}}  & T & $<$ 00:01 & $<$ 00:01 & $<$ 
			00:01 & 00:01 & 00:03 \\ 
			& O & $<$ 00:01 & $<$ 00:01 & $<$ 00:01 & 00:01 & 00:03 \\
			& D & $<$ 00:01 & $<$ 00:01 & $<$ 00:01 & $<$ 00:01 & 00:02 \\ 
			\hline
			\multirow{3}{*}{\textbf{CIFAR}} & T & 00:04 & 00:14 & 00:29 & 00:52 
			&  
			01:20 \\ 
			& O & 00:03 & 00:09 & 00:17 & 00:27 & 00:41 \\
			& D & 00:02 & 00:04 & 00:08 & 00:14 & 00:23 \\ \hline
		\end{tabular}
	\end{table}
	

\bibliographystyle{splncs04}
\bibliography{main}

\end{document}